\documentclass[letter,scriptaddress,superscriptaddress,preprint, prl]{revtex4}

\usepackage{graphicx}
\usepackage{epstopdf}
\usepackage{amsmath}
\usepackage{bm}
\usepackage{epsf,epsfig,graphics}
\usepackage{float}
\usepackage{makeidx}
\usepackage{latexsym}
\usepackage{amssymb}
\usepackage{subfigure}
\usepackage{amsmath}
\usepackage{amsfonts}
\usepackage{graphicx}
\usepackage{dcolumn}
\usepackage{amsmath}
\usepackage{color}
\usepackage{subfigure}
\usepackage{natbib}
\usepackage{booktabs}
\usepackage{multirow}
\usepackage{bm}

\begin{document}

\title{Detection of Anticipatory Dynamics Between a Pair of Zebrafish}

\author{Chun-Jen Chen, Chi-An Lin, Heng Hsu, Jos\'{e} Jiun-Shian Wu, Yu-Ting Huang and  C. K. Chan}
        
\affiliation{Institute of Physics, Academia Sinica,  Taipei, Taiwan 115, R.O.C.}

\date{\today}
\begin{abstract}
Trajectories from a pair of interacting zebrafish are used to test for the existence of anticipatory dynamics in natural systems. Anticipatory dynamics (AD) is unusual in that causal events are not necessarily ordered by their temporal order. However, their causal order can still be established if the direction of information flow (DIF) is known. In order to obtain DIF between trajectories of the two fish, we have made use of the difference of the transfer entropy between the trajectories with a history length established by experiments with known DIF. Our experimental results indicate that AD can be observed much more often in fish pairs of different genders. The use of DIF to determine causal order is further verified by the simulation of two chaotic Lorenz oscillators with anticipatory coupling; mimicking the interaction between the fish. Our simulation results further suggest that the two fish are interacting with their own internal dynamics, not by adaptation.




\end{abstract}
\maketitle


\section{Introduction}
It is our daily experience that causal events are ordered by time. One expects that effects follow their causes in time. But is this always the case? Recent advances in nonlinear physics discover that causal events are not always ordered by time when there are anticipatory interactions between the events. For example, in the phenomenon of anticipating synchronization (AS) proposed by Voss \cite{voss2000anticipating}, a slave system driven by a master can produce responses ahead of its master in time. Although there are reports on the observations of AS in various specially arranged physical systems \cite{tang2003experimental} to mimic the simulation, one wonders if this counter intuitive phenomenon can also be observed in nature. 


One of the pre-requisite for anticipatory dynamics is that systems involved must be able to process incoming information and make use of it to anticipate future events in the incoming signal. This requirement is not easily met in natural physical systems. But anticipation is an instinct for most animals and can be found in early visual system \cite{berry1999anticipation}. It is known that animals are using anticipation to compensate delays in the sensory or motor systems \cite{nijhawan2009compensating}. Anticipatory interactions might also play an important role in the collective motions as seen in the flocking of birds \cite{vicsek1995novel} or schooling of fish \cite{katz2011inferring} which are still poorly understood. Presumably, during flocking or schooling, animals do not normally run into each other because they can anticipate the future positions of their group members. Therefore, interactions between moving animals might be the best place to look for anticipatory dynamics (AD) in nature. 

In this article, we report results of our experiments designed to study anticipatory interaction between a pair of zebrafish in visual contact through their motion trajectories ($u(t)$ and $v(t)$). The method of time lag mutual information (TLMI) \cite{vastano1988information} is first used to obtain the temporal relation between $u(t)$ and $v(t)$. If the events in times series $u(t)$ is found to be ahead of the time series $v(t)$, there are two possible scenarios: either "$v(t)$ follows $u(t)$" or "$u(t)$ anticipates $v(t)"$. In order to resolve these two cases, we made use of the method of transfer entropy (TE) \cite{schreiber2000measuring} to detect the direction of information flow (DIF) between the two fish. However, the DIF is found to be sensitive to history length used in the computation of TE and additional one-way mirror experiments with known DIF are needed to establish the correct history length. Our results indicate that AD can be observed much more often in fish pairs of different genders. The validity of our anticipatory dynamics detection method of using DIF is also checked by simulations of two chaotic Lorenz oscillators with anticipatory coupling; mimicking the interaction between the fish. Our simulation results further suggest that the two fish are interacting with their own internal dynamics to perform anticipation. 


%
\section{Experiment Setup}
\begin{figure}[h!]
	\begin{center}
	\includegraphics[width=10cm]{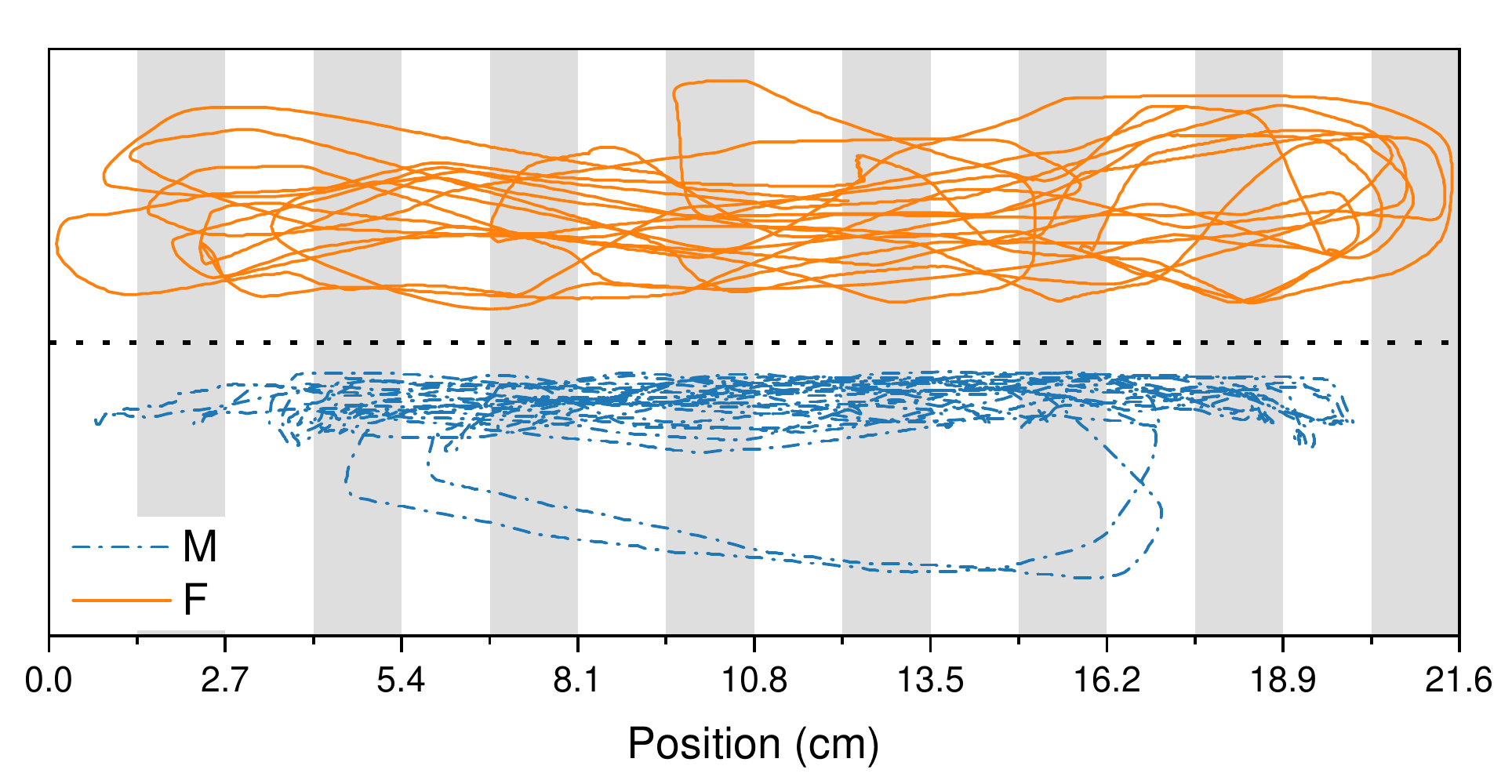}
	\end{center}
	\caption{Typical 80s trajectories of two fish in the tank. The tank was divided into two equal halves by a transparent window (the dotted line) which can be turned into a one-way mirror. These quasi-one-dimensional  trajectories were divided into 16 spatial states along the length of the tank shown as vertical strips. Different shading of the strips are used to guide the eyes. Although the two trajectories have different appearance, no correlations were found between the shapes of the trajectories with gender or role in anticipatory dynamics. Keys: M=male fish and F=female fish.}
	\label{Fig1}
\end{figure}

In the experiment, a pair of zebrafish was placed in a two-channel rectangular glass tank (Figure~\ref{Fig1}) with the two fish separated by a window such that the two fish were visible to each other. The rectangular channels were of size $21.6 cm \times 4.5 cm$. The fish were 1 $\sim$ 1.5 years old and of lengths $3.5 \sim 4  cm$.  The water levels in the two channels were kept around 2 cm deep. This depth was determined empirically to be deep enough to avoid the jumping of fish out of the water but not so deep that we can treat the fish as being confined in a quasi-one-dimensional channel. The window between the two channels was made of Plexi-glass and it can be turned into a one-way mirror when a semi-reflective filter is attached to it together with different illumination levels in the two channels. The tank was placed on top of a LCD monitor ($ 40 cm \times 70 cm$) which was controlled by a computer to provide different spatially uniform illuminations for the two channels. When the one-way mirror configuration is enabled, only the fish in the dark channel can see the other fish in the bright channel while the fish in the bright channel can only see its own reflections from the walls of the channel. A CCD camera (Basler acA4096-40um) placed 90 cm above the tank was used to capture images of the fish. An IR LED ($940 nm$) linear array was also used to provide illumination which was visible only to the CCD but not to the fish so that captured image quality can still be maintained even when visible illumination levels in the two channels were low. The whole setup was kept inside a box with black walls ; preventing the fish from detecting spacial visual cues. 

A total of more than $20$ fish were kept in a large tank and selected fish were then placed into the two channels for experiments. Experiments were performed from two groups (6 each: 3 males and 3 females) which were obtained at different dates. We found that a settlement time of 0, 2.5, 5 and 10 minutes before the start of experiment recording did not affect the results of our experiments. Trajectories of the two fish during experiments were obtained from video images recorded at 30 or 60 frames per second (fps) with a spatial resolution of $1340 \times 560$ square pixels covering an area of $21.6 cm \times 9 cm$. The recording duration is always set to 500 s. For the health of the fish, the total experiment period per day was limited to 60 minutes. Trajectories of the fish were extracted from recorded images by the scikit-image packages.  All experiments were performed at $25 ^\circ C$. Data reported below were obtained from experiments recorded mainly at 30 fps as data from 60 fps recording give similar results. Figure~\ref{Fig1} shows trajectories of two fish separated by a transparent window in the tank during a typical experiment. The horizontal span of the tank was divided into 16 discrete locations and the quasi-one-dimensional trajectories are represented by these 16 states for the computation of TLMI and TE below.

\section{Experimental Results}
\begin{figure}[h!]
	\begin{center}
		\includegraphics[width=5.4cm]{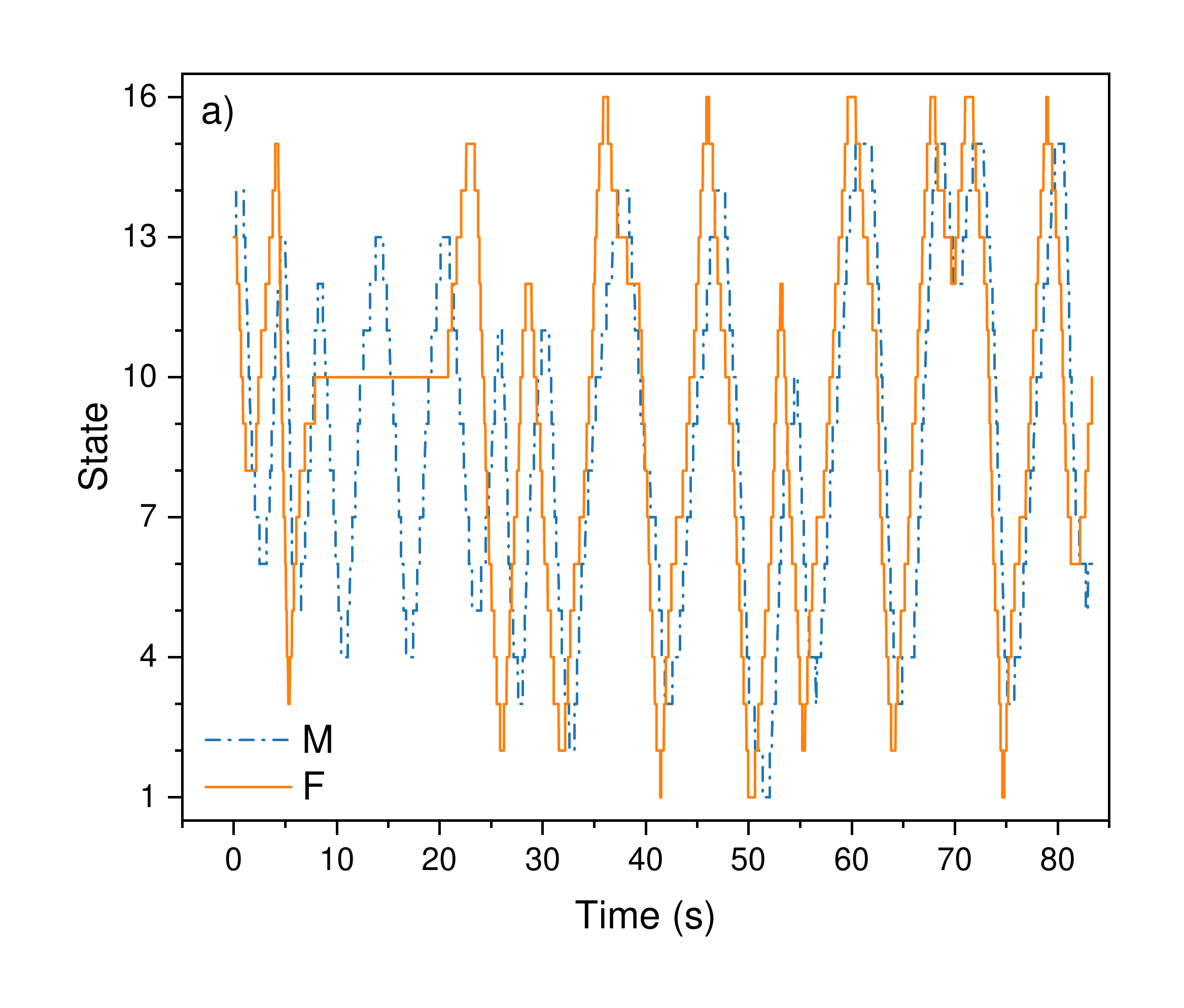}
		\includegraphics[width=5.4cm]{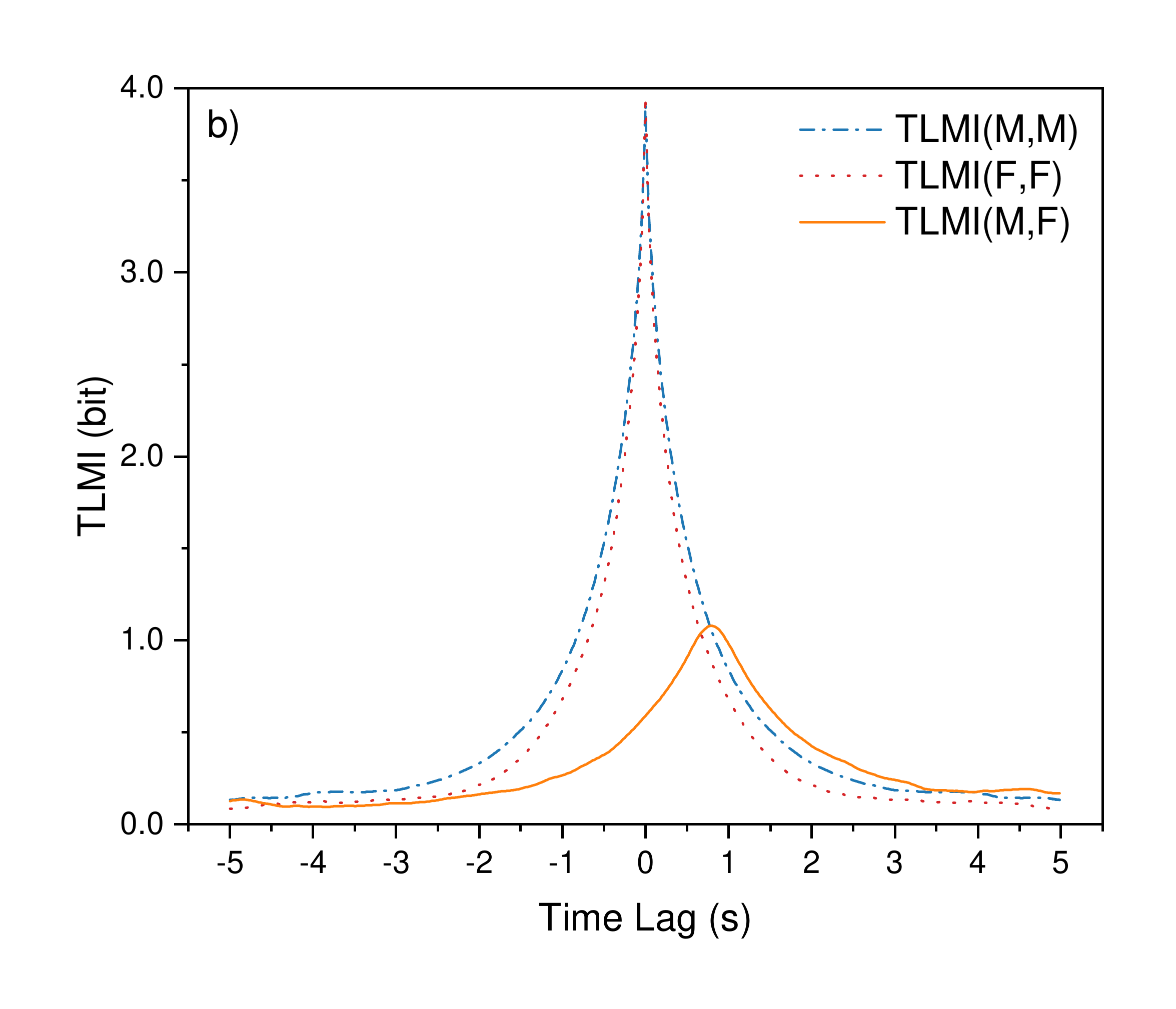}
		\includegraphics[width=5.4cm]{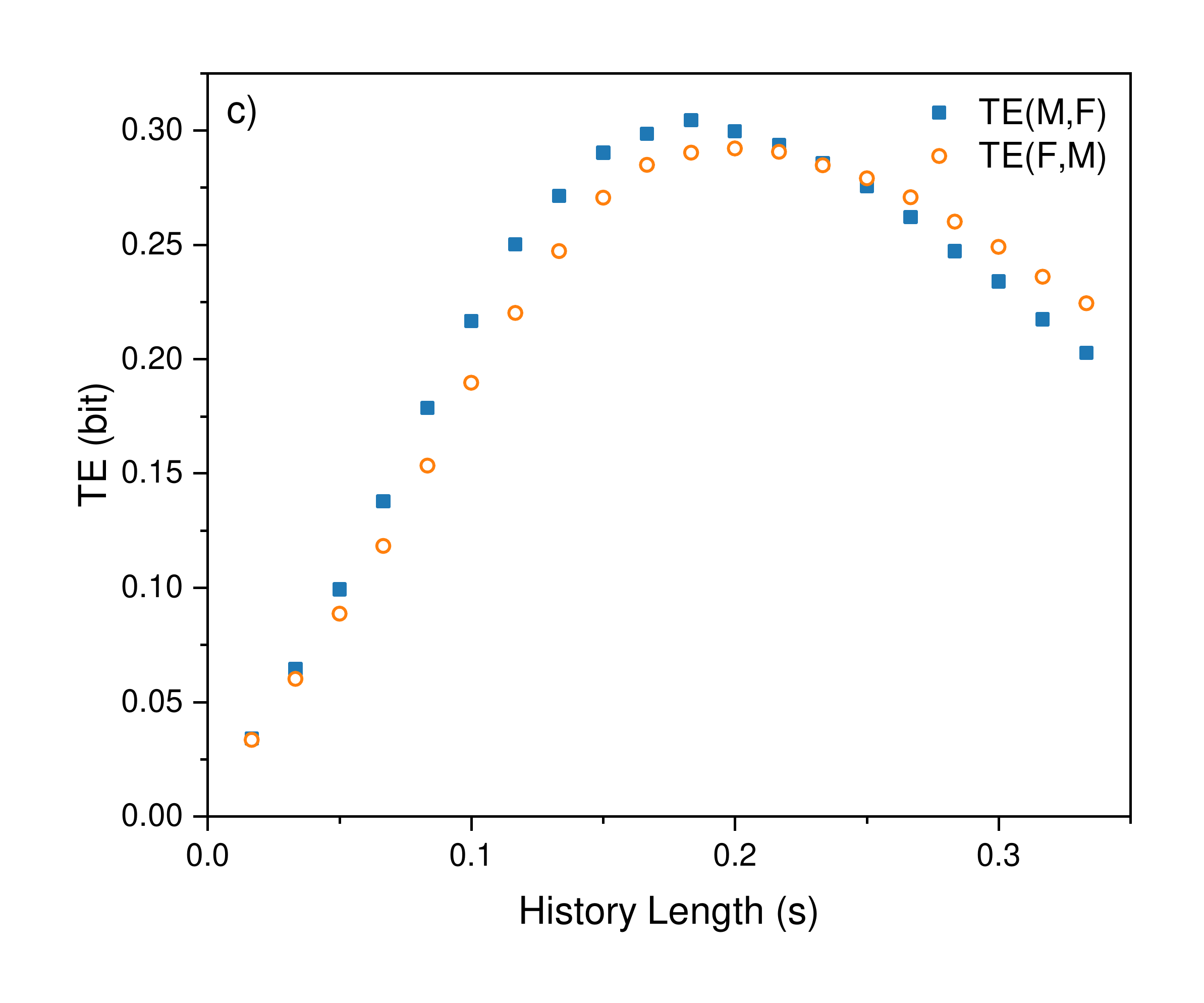}
	\end{center}
	\caption{Typical experimental results for an interacting pair of fish of different genders. a) The time course of the states from the trajectories of Figure~\ref{Fig1}. It can be seen that F is leading M. b) TLMI as a function of time lag computed from the two time series shown in a). c) History length dependence of TE for both directions from $F \rightarrow M $ and $M \rightarrow F$. Keys: M=male, F=female, TLMI(A,B) = time lag mutual information between A and B and TE(A,B) = transfer entropy from A to B.}
	\label{Fig2}
\end{figure}

Figure~\ref{Fig2} is the results of the analysis of the trajectories shown in Figure~\ref{Fig1}. The measured trajectories of fish were represented as two time series: $U \equiv \{u_i\}$ and $V \equiv \{v_i\}$ shown in Figure~\ref{Fig2}a. Here $u_i$ and $v_i$ are the positions (discretized in to 16 states) of the two fish respectively at the $i^{th}$ step with a step size of $1/60s$. It can be seen that the two fish were always close to each other; indicating that there was an attractive interaction between them. We found that the interaction between the two fish was visual. There was little interaction between the two fish when the background illumination provided by the LCD monitor was switched off. The background illumination in Figure~\ref{Fig2} was 135 $cd/m^2$. The interaction was found to be not too sensitive to this background illumination; similar results can be obtained when the background illumination was changed by $\pm 30\%$.

We have also performed experiments similar to that shown in Figure~\ref{Fig1} in tanks with different lengths. For a longer tank (length = 44cm), since the two fish can be quite far apart, there were interactions between them only when they were visible and close to each other; resulting in a much smaller peak in the measured cross-TLMI curve (Figure~\ref{Fig2}b). On the other hand, for a shorter tank (length = 10cm), it seems that the fish did not adapt well to such a confined space and they can be seen to move back and forth between the two ends of the channels endlessly and not interested in each other; resulting in a TLMI curve with no peak. Our choice of a tank with a length of 21.6cm was based on these observations to optimize the interactions between the two fish.

To investigate the temporal relation between $U$ and $V$, we make use of the TLMI \cite{chen2017characterization} between $U$ and $V$ which is defined as:

\begin{equation}
TLMI(U,V,\delta t) = \sum_{u_{t+\delta t}, v_t} P(u_{t +\delta t},v_t)log[\dfrac{P(u_{t+\delta t},v_t)}{P(u_{t+\delta
t})P(v_t)}]
\end{equation}

\noindent
where $P(...)$ is the probability distribution or joint distribution of the variables in $(...)$.
It measures how much information is being shared between $\{u_i\}$ and $\{v_i\}$ when $U$ is shifted $j$ step ($\delta t = j\Delta t$) ahead of $V$ with $\Delta t$ being the step size. The TLMI from the same time series is also called the auto-TLMI and cross-TLMI for those from two different time series. The TLMI between $U$ and $V$ as a function of  $\delta t$ (time lag) is shown in Figure~\ref{Fig2}b. As references, the figure also shows the corresponding auto-TLMI of $U$ and $V$ which measure how information between two temporal points ($\delta t$ apart) in the same time series are being shared. The peak position of $TLMI(U,V,\delta t)$ provides information of how events in $U$ and $V$ are related temporally \cite{chen2017characterization}.  The peaks for the auto-TLMI are of course located at $\delta t = 0$. It can be seen that both $TLMI(U,U,\delta t)$ and $TLMI(V,V,\delta t)$ decrease similarly in both the positive and negative $\delta t$ direction.

\begin{figure}[h!]
	\begin{center}
		\includegraphics[width=7cm]{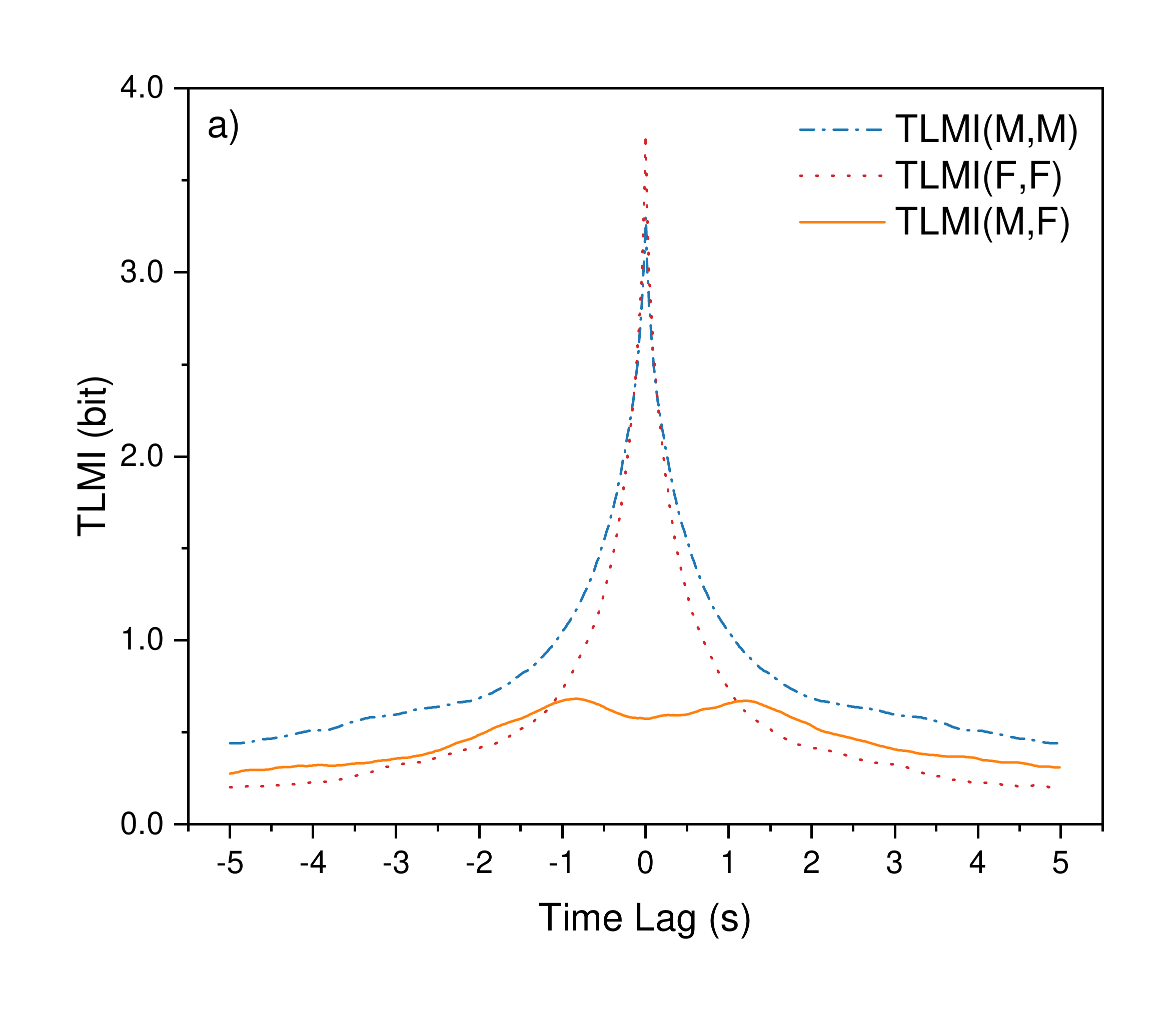}
		\includegraphics[width=7cm]{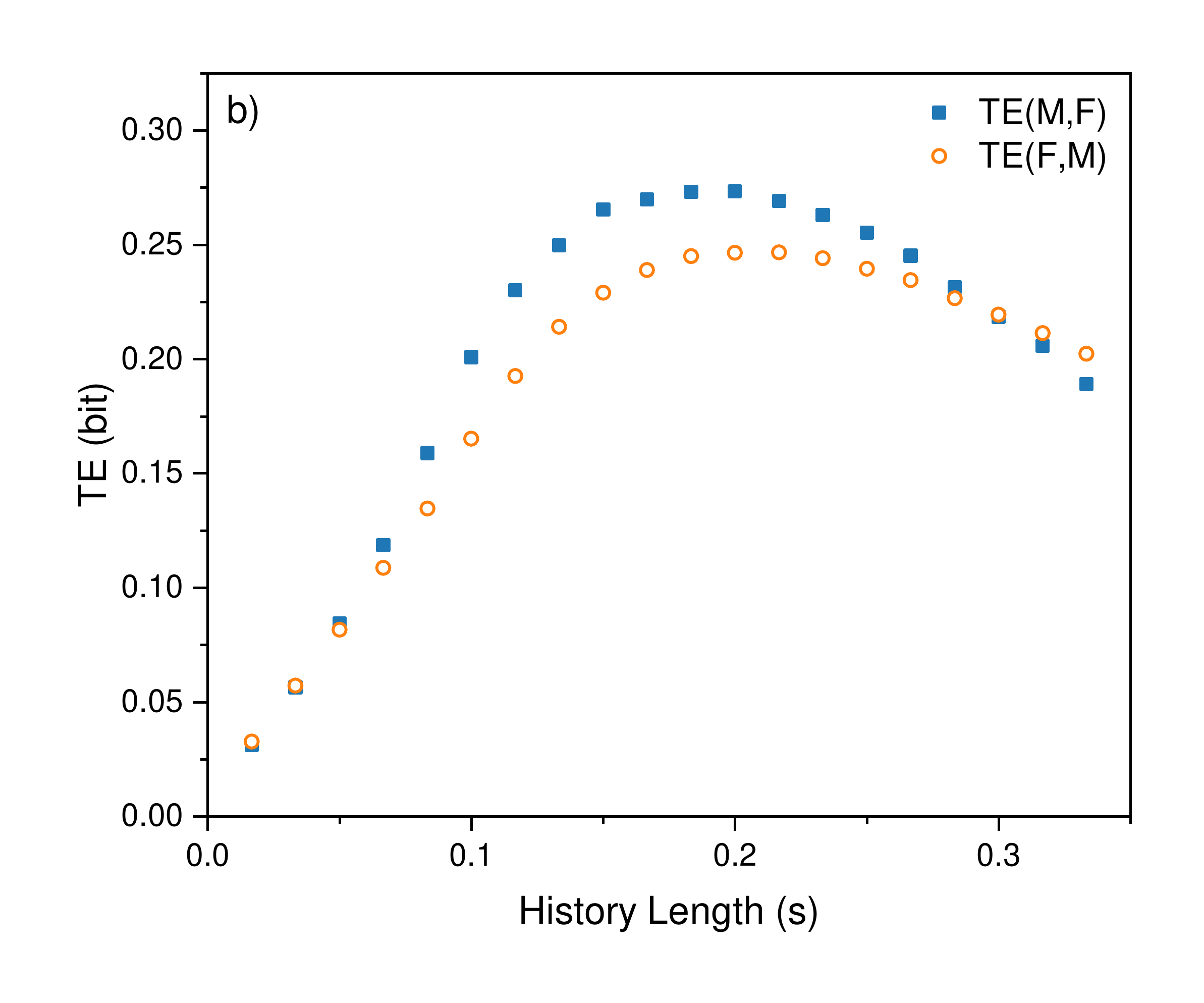}
	\end{center}
	\caption{A typical example of 2-peak TLMI in experiments similar to that shown in Figure~\ref{Fig2}. Two peaks can be clearly seen in the measured TLMI. These peaks could origin from a scenario that the male and the female fish can be leading at different periods of time during the experiment.}
	\label{Fig3}
\end{figure}

Different from the auto-TLMI, the peak of the cross-TLMI between $U$ and $V$ can be located either at $\delta t > 0$ or $\delta t < 0$, depending on the temporal relation between $U$ and $V$. A peak at $\delta t = \delta t_p > 0$ for $TLMI(U,V,\delta t)$ indicates that $V$ at $t-\delta t_p$ contains most information of $U$ at $t$. In other words, events in $V$ is leading events in $U$. This peak location is also called the anticipatory horizon. This last statement could mean one of two possibilities: either a) $U$ follows $V$ or b) $V$ anticipates $U$ and vice versa for peaks at $\delta t<0$. However, in some experiments, two peaks in the TLMI can be observed as shown in Figure~\ref{Fig3}. In such 2-peak TLMI cases, one possible interpretation is that events in $V$ are leading those in $U$ only during some of the time and vice versa for events in $U$. From our experiments, there is only a single peak in the majority of the measured TLMI (see Table~\ref{Summary}).


For the cases of TLMI with a single peak, we made use of the method of TE to compute the direction of the net flow of information between the two fish to resolve the two possibilities of "anticipate" and "follow". We follow the definition of TE of Schreiber \cite{schreiber2000measuring}, and calculate the TE from time series $U$ to $V$ as:
\begin{equation}
TE(U,V)(k, l, t)=H(V(t)|V^{(k)}(t-1)) - H(V(t)|V^{(k)}(t-1);U^{(l)}(t))
\end{equation}
where $H(.|.)$ denotes the conditional entropy. The notation $V^{(k)}(t)=\{v_{t-k},v_{t-k+1},...,v_{t-1}\}$ represents the history of $V(t)\equiv v_t$ with a length of $k$ time step before the time bin $t$ and similarly for $U$. This TE measures the reduction of entropy of the $V(t)$  by using both $V$'s own history and that of $U$ with history length $k$ and $l$ time step respectively. With this definition, $TE \geq 0$. If $TE(U,V)$ is positive, we will have information transfer from $U$ to $V$. Since there might be information flow in both directions, we can obtain the DIF from the net information flow between $U$ and $V$. If $TE(U,V) > TE(V,U)$, DIF is from $U$ to $V$. Usually, TE is defined at equal time for both $U$ and $V$. However, to take into account the effect of time shifted in anticipatory dynamics, we also need to shift one of the time series by $\delta t_p$ obtained in Figure~\ref{Fig2}b when we compute TE. The use of TE to detect causal relationship between two time series is not new. Butail et al \cite{butail2016model} used TE to detect "leaders" in interaction between zebrafish. However, in Ref\cite{butail2016model}, a history length of one was used and no effects of time lag due to anticipatory dynamics were considered. To simplify analysis, we have set the history lengths in both $U$ and $V$ to be the same and labeled it as $h$. Since it is not known a priori the correct history length to be used to compute the two TEs, we have computed the history length dependence of TE as shown in Figure~\ref{Fig2}c. It can be seen that the DIF can be reversed if different history lengths ($h$) were used.
 Note that the difference of TE in the two directions are quite small. Presumably, this is due to the similarity of the two trajectories.

\begin{figure}[h!]
	\begin{center}
		\includegraphics[width=5.4cm]{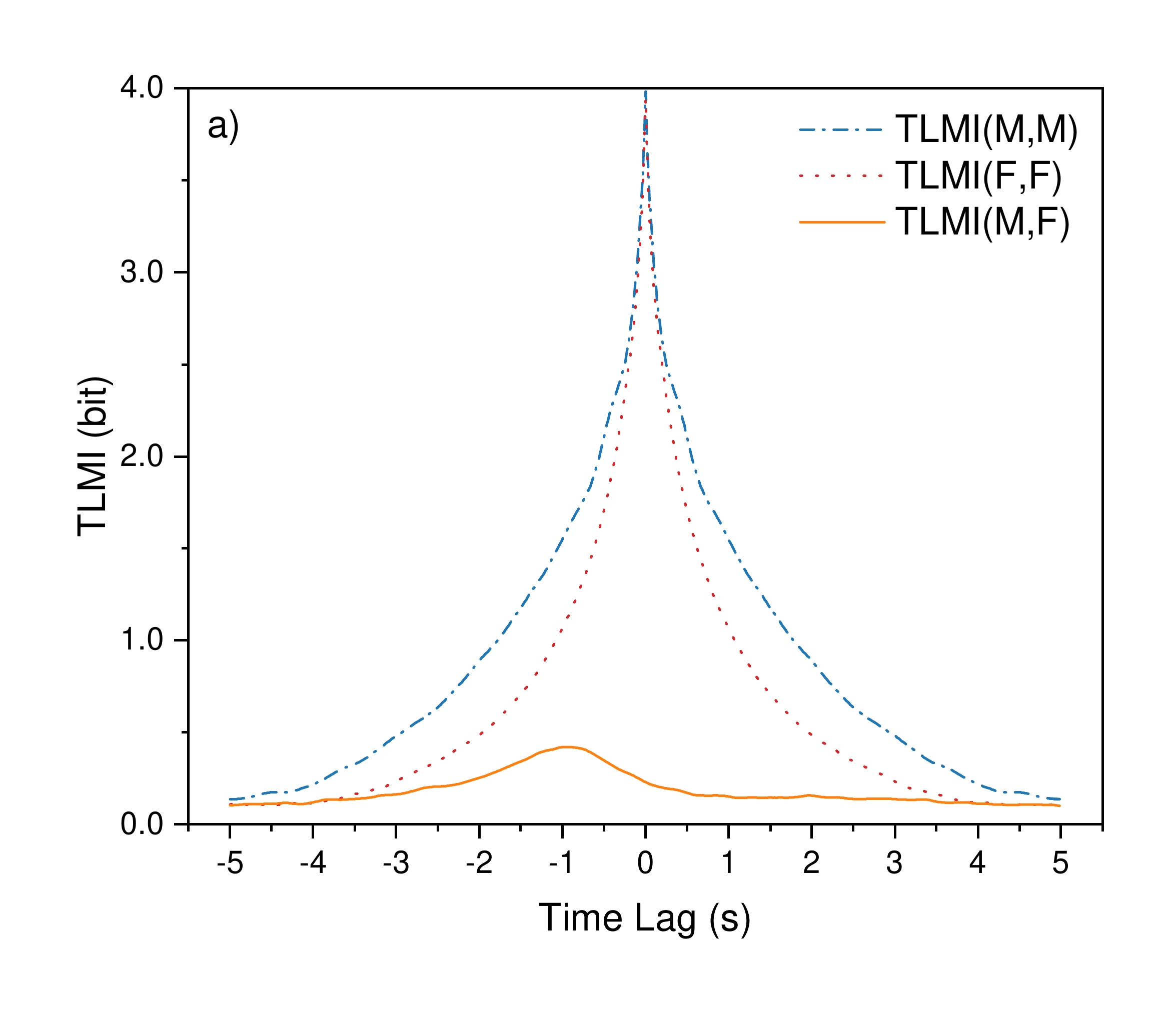}
		\includegraphics[width=5.4cm]{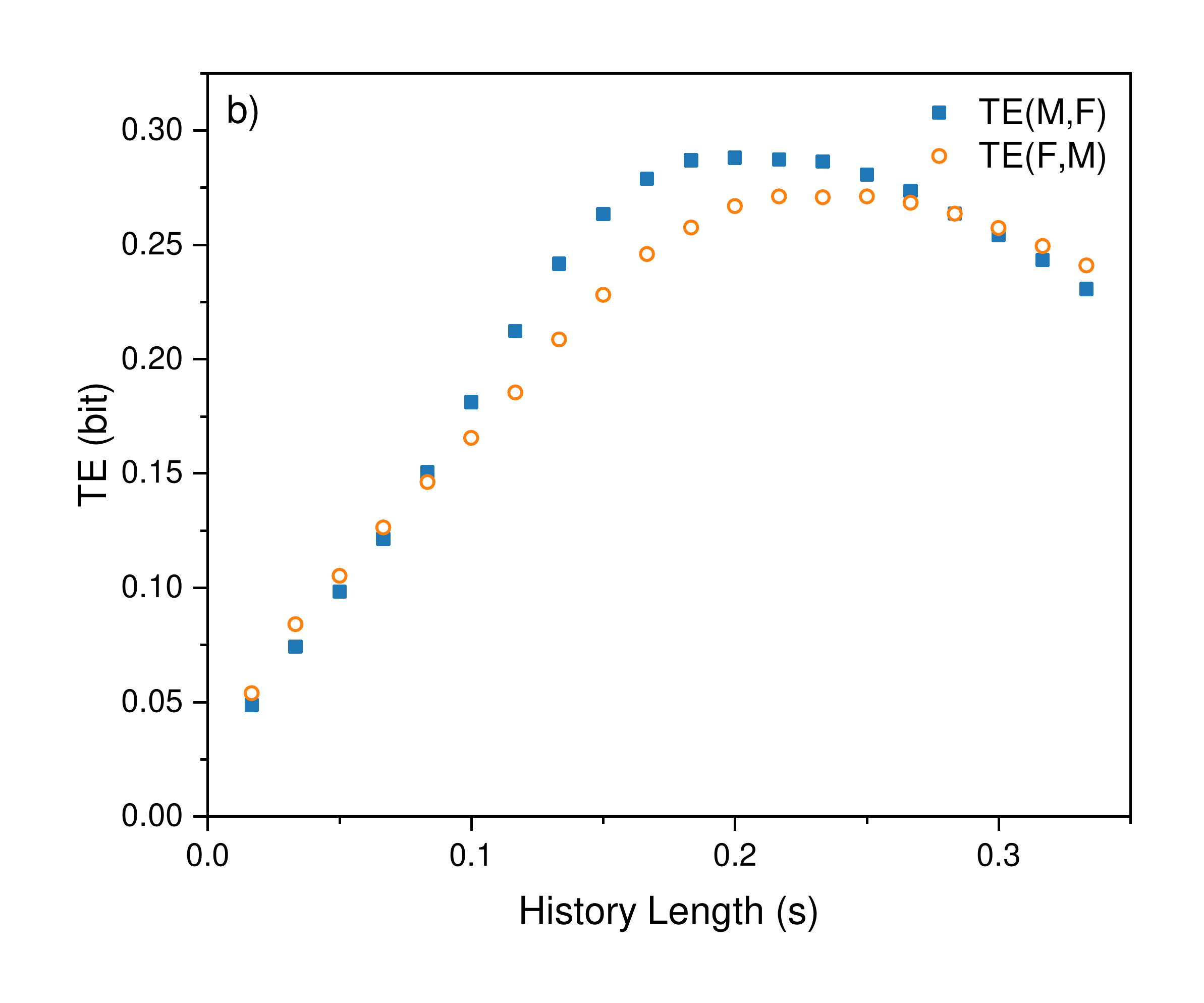}
		\includegraphics[width=5.4cm]{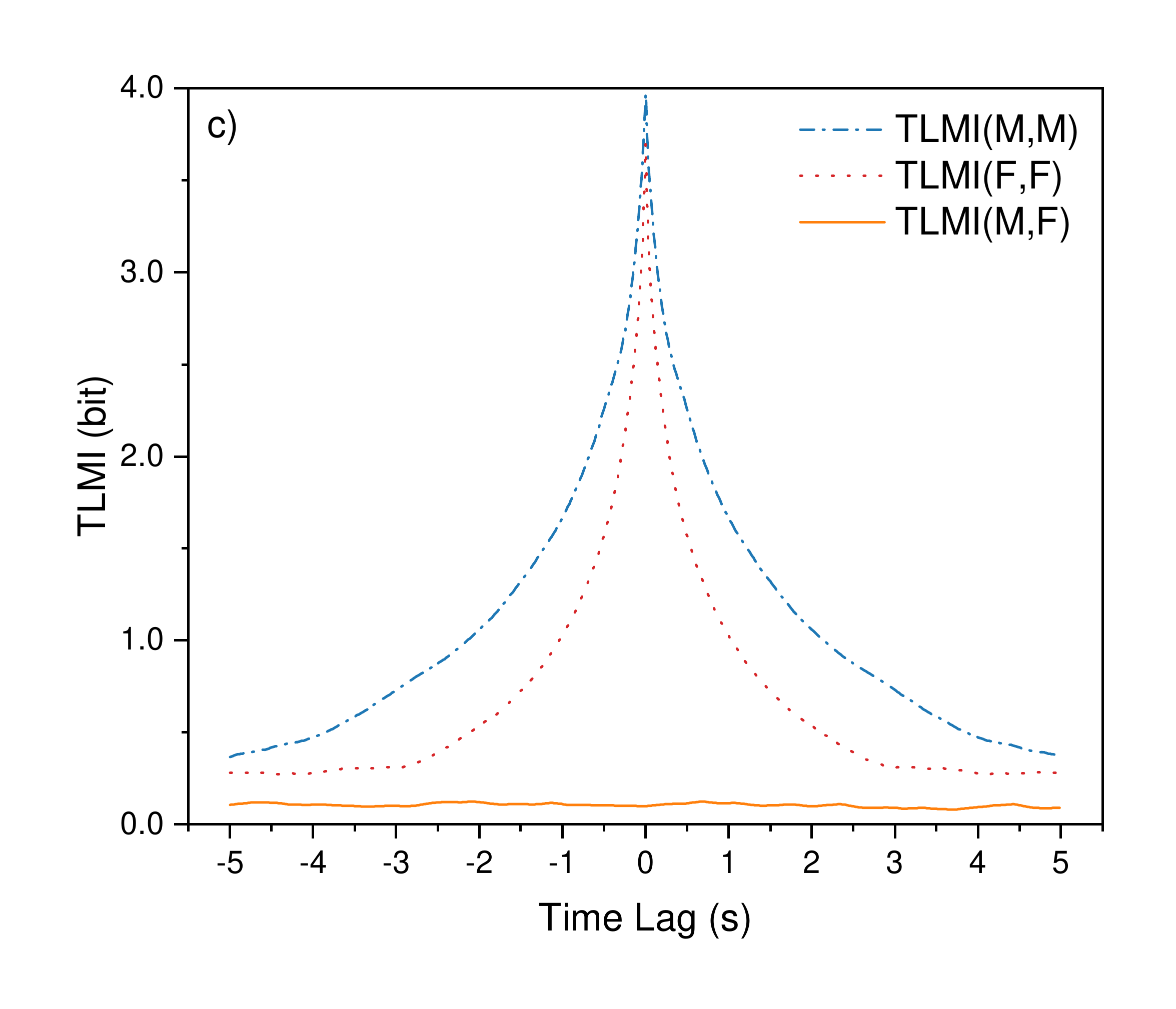}
	\end{center}
	\caption{Typical results from one-way mirror experiments: a) Time lag dependence of TLMI between Fish B in the bright and Fish D in the dark channels; showing that the trajectory of Fish B (male) is ahead of that from Fish D (female). b) History length dependence of TE for both transfer directions. Since it is known that the DIF is from the bright to the dark channel, the correct history length should be between $0.1$ and  $0.25$ s. c) The same as a) except that the two fish were exchanged. The experimental conditions here were the same as those in Figure~\ref{Fig2}; except that the transparent window between the two channels was replaced by an one-way mirror.}
	\label{Fig4}
\end{figure}
In order to determine the correct $h$ to be used, we have setup one-way mirror experiments in which only one fish can see the other and therefore the DIF is known. These one-way mirror experiments were implemented by using different illumination levels in the two channels and replacing the transparent window separating the two channels in the tank by a semi-reflective window.
Figure~\ref{Fig4} is the typical result of a one-way mirror experiment. The light intensity in the two channels were 135 $cd/m^2$ and less than 1 $cd/m^2$ for the bright and dark channel respectively. If one labels the fish in the bright/dark channel as Fish B/D respectively, then only Fish D can see Fish B while Fish B sees its own reflection in the one-way mirror as mentioned above. Therefore, information is flowing from the bright channel to the dark channel. Figure~\ref{Fig4} shows typical results from such one-way mirror experiments. Figure~\ref{Fig4}a shows that the female tended to follow the male fish, and the TE computation (Figure~\ref{Fig4}b) shows the correct direction of information flow when the history length is between $0.1$ and $0.25$ s. If we exchange the two fish in the channels, there were little interaction (Figure~\ref{Fig4}c). Also, the peak of the TLMI between Fish B and Fish D in Figure~\ref{Fig4}a was found to be only a fraction of that from Figure~\ref{Fig2}b; indicating that the interaction between the two fish in the one-way mirror experiment was weaker than that when both fish can see each other.

More one-way experiments ($N>15$) have also been carried out with different pairs of fish similar to those in regular experiments to check for the correct $h$ needed and we find that the obtained DIF will be correct if $h$ is between $0.1$ and $0.25$ s. When this calibrated criterion on $h$ was used to obtain the DIF for different regular experiments with a transparent window, we found that anticipatory dynamics can be observed mostly when the two fish are of different genders with information flowing from the males to females while the trajectories of females seemed to be leading those from the males. Table \ref{Summary} is a summary of our finding in the regular experiments (same illumination levels in both channels) based on a history length between $0.1$ and $0.25$ s. The important features from the table are: a) peaks of the cross-TLMI are higher from fish pairs of different genders, b) For pairs of different genders, we observed more cases of "anticipate" than "follow", c) females were more likely to be the net information receiver. However, point c) might be related to the fact that experiments were being performed with a certain group of fish.



\begin{table}[]
\begin{tabular}{|c|c|c|c|cc|cc|c|}
\hline
\multirow{2}{*}{}   & \multirow{2}{*}{\# of Expt} & \multirow{2}{*}{NI} & \multirow{2}{*}{2 Peaks} & \multicolumn{2}{c|}{Anticipate}      & \multicolumn{2}{c|}{Follow}         & \multirow{2}{*}{Avg MI Peak} \\ \cline{5-8}
                    &                             &                            &                          & \multicolumn{1}{c|}{FM}     & MF     & \multicolumn{1}{c|}{FM}     & MF    &                              \\ \hline
\multirow{2}{*}{F/M} & 32                          & 5                          & 7                        & \multicolumn{1}{c|}{11}     & 4      & \multicolumn{1}{c|}{4}      & 1     & \multirow{2}{*}{20.3\%}      \\ \cline{2-8}
                    & 100\%                       & 15.6\%                     & 21.9\%                   & \multicolumn{1}{c|}{34.4\%} & 12.5\% & \multicolumn{1}{c|}{12.5\%} & 3.1\% &                              \\ \hline
\multirow{2}{*}{F/F} & 18                          & 5                          & 5                        & \multicolumn{2}{c|}{3}               & \multicolumn{2}{c|}{5}              & \multirow{2}{*}{13.5\%}      \\ \cline{2-8}
                    & 100\%                       & 27.8\%                     & 27.8\%                   & \multicolumn{2}{c|}{16.7\%}          & \multicolumn{2}{c|}{27.8\%}         &                              \\ \hline
\multirow{2}{*}{M/M} & 18                          & 10                         & 0                        & \multicolumn{2}{c|}{6}               & \multicolumn{2}{c|}{2}              & \multirow{2}{*}{11.6\%}      \\ \cline{2-8}
                    & 100\%                       & 55.6\%                     & 0.0\%                    & \multicolumn{2}{c|}{33.3\%}          & \multicolumn{2}{c|}{11.1\%}         &                              \\ \hline
\end{tabular}
	\caption{Summary of experimental results of detecting anticipatory interaction between a pair of zebrafish with same or different genders based on the measurement of cross-TLMI and difference of TE with history length between $0.1$ and $0.25s$. There were three different genders combination in the experiments: F/M - female and male fish in the two channels and similarly for other two cases of F/F and M/M. Results of the experiments are classified as NI (No Interaction), 2 Peaks, Anticipate and Follow. For both the "Anticipate" and "Follow" conditions, the cases are further subdivided into FM (female anticipates/follows male) and MF (male anticipates/follows female). About half of the cases from different genders data show anticipation and these pairs are interacting better than those of same gender as shown in the column "Avg MI Peak" (the average cross-TLMI peak height measured as the percentage of the auto-TLMI peak height).}
	\label{Summary}
\end{table}


\section{Comparisons with Lorenz oscillators}

Our method of using difference of TE to determine the direction of information flow has also been tested by the simulation of two Lorenz attractors\cite{pecora1990synchronization} ($L_1$ and $L_2$) with anticipatory coupling through their $x$ variables. We used the anticipatory coupling mechanism introduced by Voss \cite{voss2000anticipating}. The governing equations for the $L_1$ are:

\begin{eqnarray}
{dx_1\over{dt}} = \sigma(y_1-x_1)+k_1[x_2(t)-x_1(t-\tau)] +\lambda_1\eta_1(t));\\
{dy_1\over{dt}} = x_1(\rho -z_1) - y_1;
{dz_1\over{dt}} = x_1y_1 - \beta z_1
\end{eqnarray}

\noindent
Similarly for $L_2$:

\begin{eqnarray}
{dx_2\over{dt}} = \sigma(y_2-x_2)+k_2[x_1(t)-x_2(t-\tau)] +\lambda_2\eta_2(t));\\
{dy_2\over{dt}} = x_2(\rho -z_2) - y_2;
{dz_2\over{dt}} = x_2y_2 - \beta z_2
\end{eqnarray}

\noindent
The anticipatory interaction in $L_1$ is the term $k_1[x_2(t)-x_1(t-\tau)]$ where $\tau$ is related to anticipatory horizon of $L_1$ with respect to $L_2$ and $k_1$ the interaction strength. A single $\tau$ is used here for simplicity to simulate the situation where both fish have similar anticipatory capabilities. However, $k_1$ and $k_2$ can be different because different fish or fish of different genders could have different roles during interaction. The usual Lorentz parameters $\sigma$, $\beta$ and $\rho$ are 10, 8/3 and 28 respectively so that the system is in the chaotic regime. The interaction between $L_1$ and $L_2$ is also made noisy by the noise term with $\lambda_1$ and $\lambda_2 $ being the amplitude of the Gaussian noise ($\eta_1$ and $\eta_2$) with zero mean and unity standard deviation. The noise term is needed in order to produce results similar to experiments as will be shown below. All the parameters here are dimensionless. Euler's method is used to integrate the models with a time step $\Delta t = 0.001$ and a smaller time step will not change the essential features of the results. The delayed feedback time $\tau$ is always set to 0.05 for all the simulation. The values of $k_1$, $k_2$, $\lambda_1$, and $\lambda_2$ reported below are chosen to reproduce TLMI similar to those observed in experiments for various conditions. 
\begin{figure}[h!]
	\begin{center}
		\includegraphics[width=5.4cm]{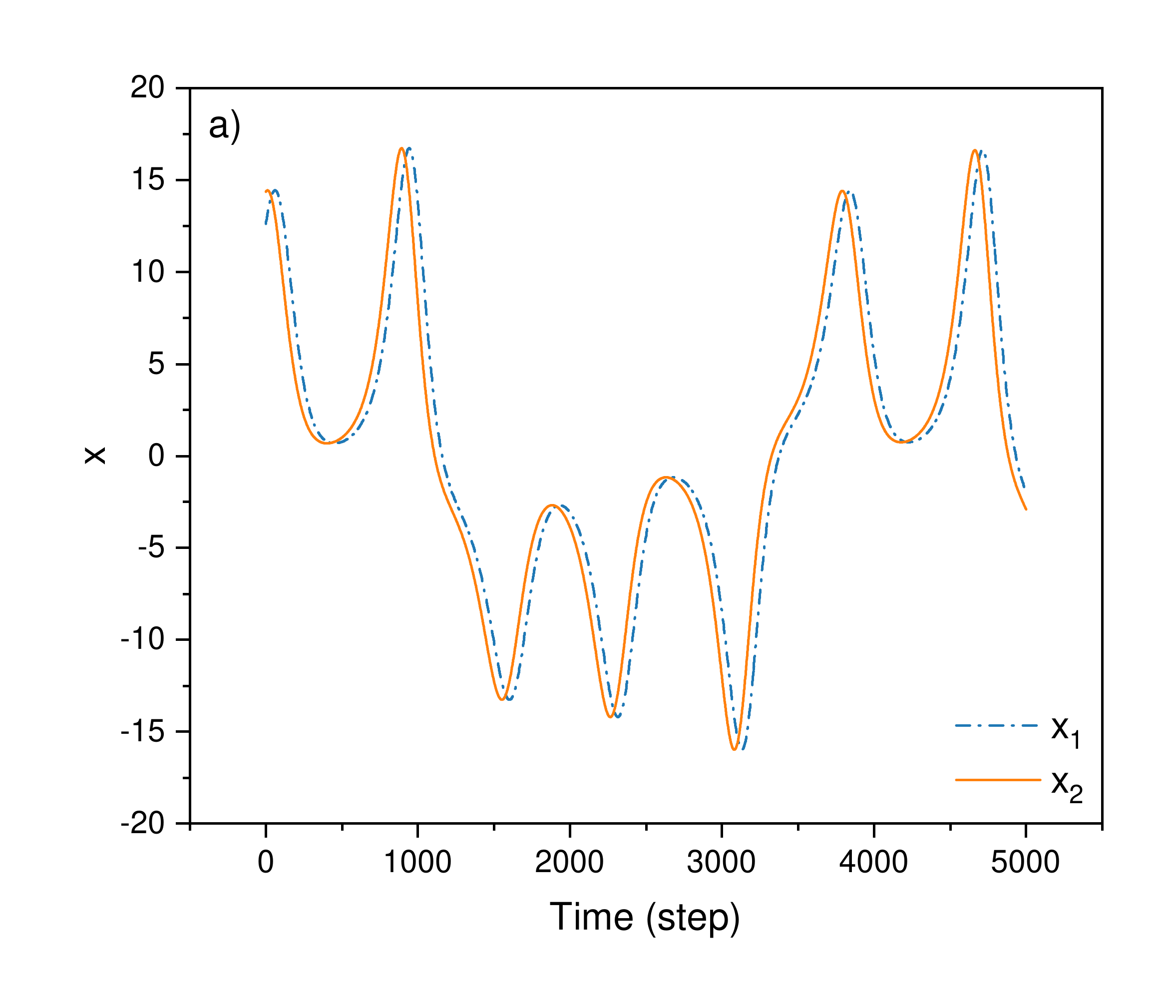}
		\includegraphics[width=5.4cm]{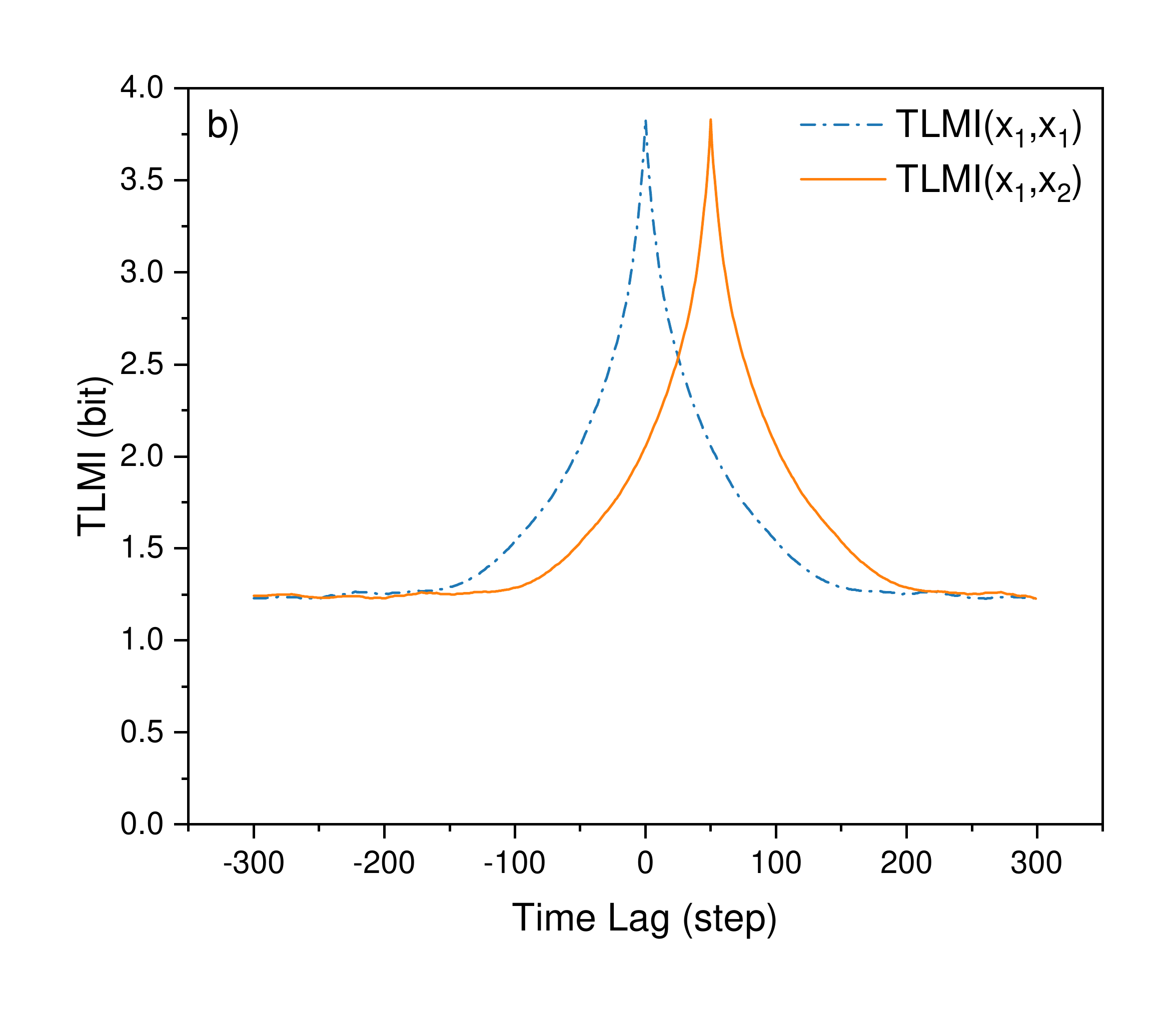}
	\end{center}
	\caption{Simulation results of two chaotic Lorenz oscillators ($L_1$ and $L_2$ ) coupled through their $x$ variables; with $L_1$ being the master and $L_2$ being the slave. a) The time course of the $x_1$ and $x_2$; showing the leading of $x_2$.  b) The TLMI between  $x_1$ and $x_2$; showing that either "$L_1$ is anticipating $L_2$" or "$L_2$ is following $L_1$". c) The history length dependence of TE from $x_1$ to $x_2$ and vice versa. The simulation parameters are: $\Delta t = 0.001$, $\tau = 0.05$, $k_1=0, k_2=1, \lambda_1=0$ and $\lambda_2=0$.\textcolor{red}{}}
	\label{Fig5}
\end{figure}
\subsection{Anticipation in a Master-Slave system without noise}
We will consider the simple case without noise first. If $k_1$ is zero, $L_1$ is autonomous and acts like a master while $L_2$ is a slave which can only receive information from $L_1$. The system $L_2$ will synchronize with $L_1$ when the anticipatory interaction $k_2[x_1(t)-x_2(t-\tau)]$ vanishes. Therefore, one would expect $L_1$ to be in a chaotic state while $L_2$ is synchronized with $L_1$ with $x_2(t-\tau) \sim x_1(t)$, $y_2(t-\tau) \sim y_1(t)$ and $z_2(t-\tau) \sim z_1(t)$ for $k_1 = 0$ with a suitable $k_2$. Figure \ref{Fig5} is a typical result of simulations for such a master slave system ($k_1 = 0$) with the noises in both $L_1$ and $L_2$ set to zero. It can be seen that the two systems can be synchronized (Figure~\ref{Fig5}a) by the coupling only in the $x$ variable  with $L_2$ being ahead of $L_1$. In other words, $L_2$ is anticipatory of $L_1$. However the corresponding TLMI shown in Figure~\ref{Fig5}b is very different from those found in experiments. The peak of the cross-TLMI has the same height as that of the auto-TLMI. The equal heights of the peaks can be understood from fact that the two system are synchronized with a phase (time) shift and therefore the cross-TLMI should be the same as the auto-TLMI except with only a shift in time (time lag). No history dependence of TE is shown because our simulation results show that TE in both direction are always equal as expected. Note that the $k_2 $ and $\tau$ are chosen so that the two systems can synchronize. If $k_2$ or $\tau$ is too large, the two oscillators cannot be synchronized. The phenomenon reported here is known as anticipating synchronization \cite{voss2000anticipating}. 

\subsection{Anticipation in a Master-Slave system with noise}
\begin{figure}[h!]
 	\begin{center}
 		\includegraphics[width=5.4cm]{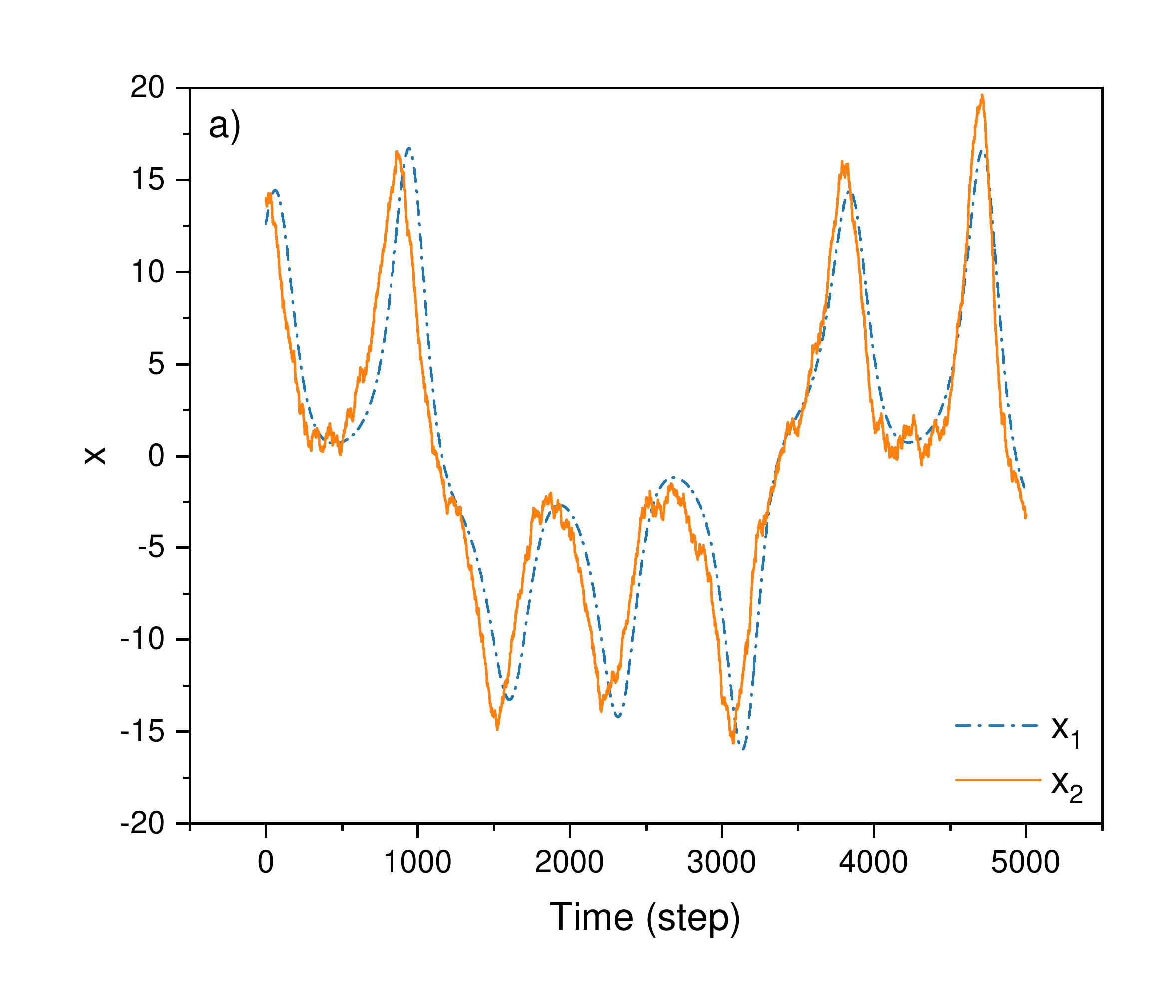}
 		\includegraphics[width=5.4cm]{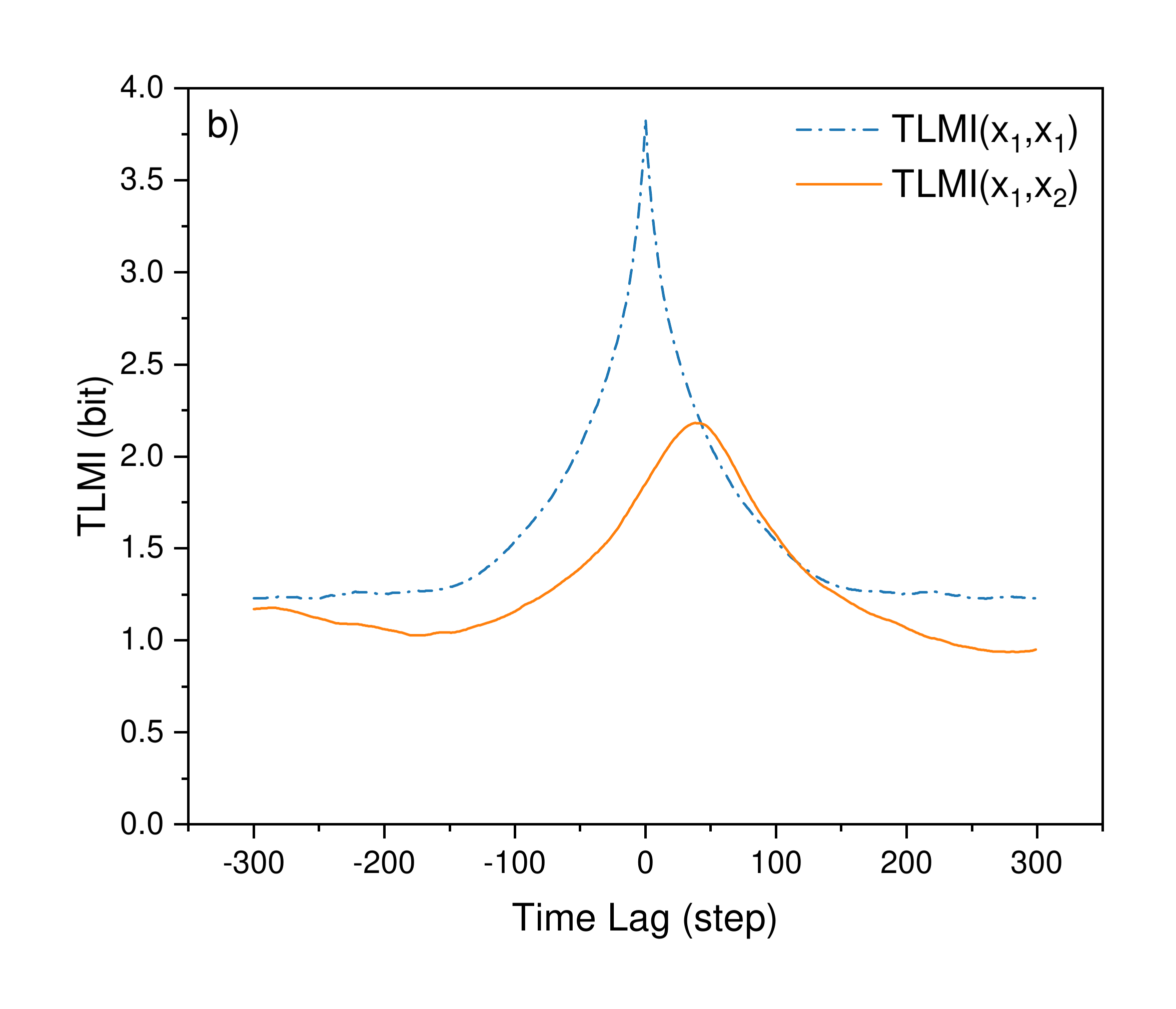}
 		\includegraphics[width=5.4cm]{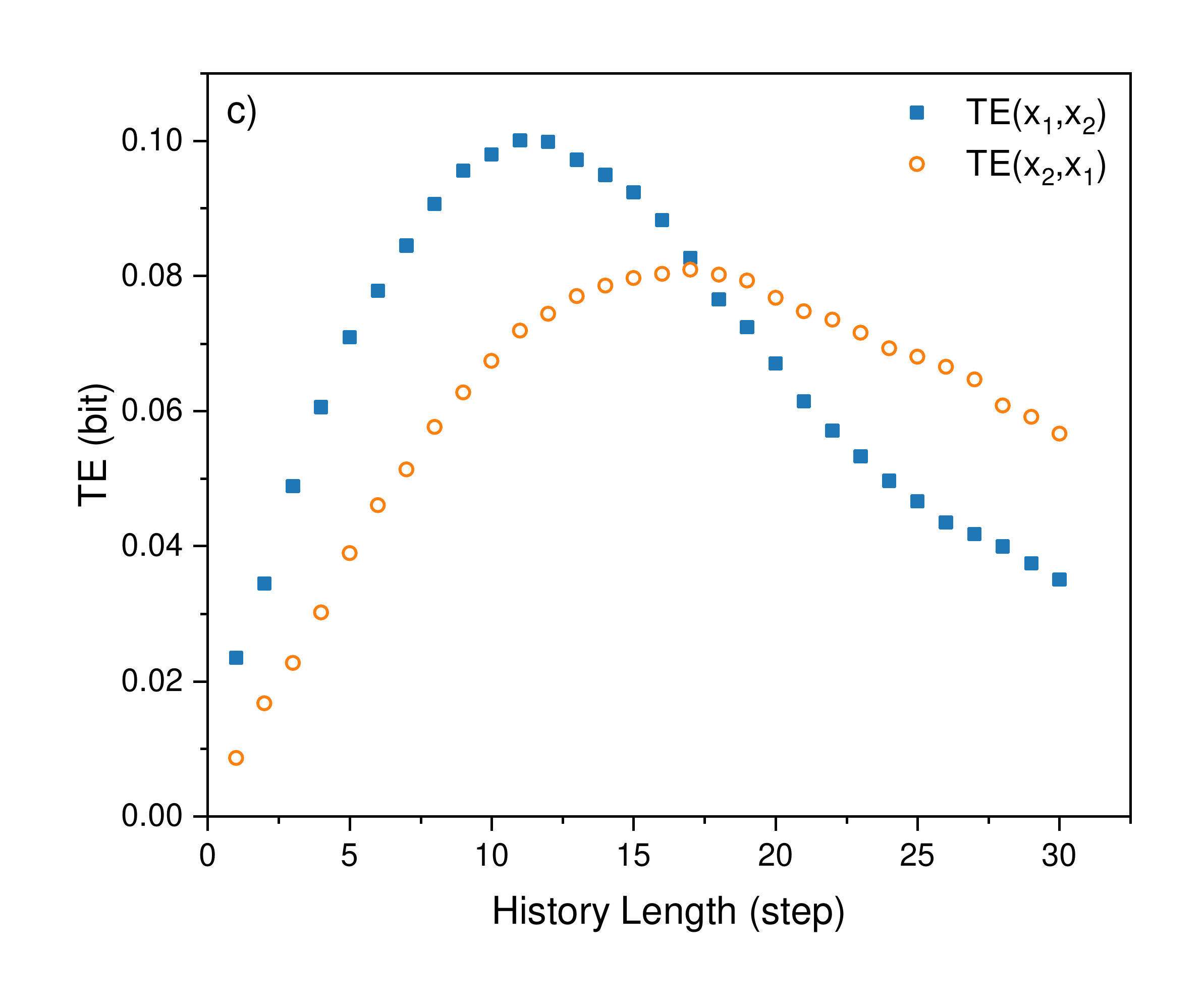}
 		
 	\end{center}
 	\caption{Simulation results of $L_1$ and $L_2$ with noisy ($\lambda_1 =0$ and $\lambda_2 = 3.8$) anticipatory interaction ($\tau = 0.05$); with $L_1$ being the master ($k_1=0$) and $L_2$ being the slave ($k_2=1$). a) The time course of the $x_1$ and $x_2$; showing the leading of $L_2$.  b) The TLMI between  $x_1$ and $x_2$. c) The history length dependence of TE from $x_1$ to $x_2$ and vice versa. The direction of information flow is correct only when the history length is taken between 1 and 17 simulation time step.}
 	\label{Fig6}
 \end{figure}

Figure~\ref{Fig6} shows typical results of a master-slave ($k_1=0$) system when $\lambda_1 =0 $ and $\lambda_2 \neq 0$. Similar results can also be obtained when both $\lambda$ are not zero (not shown). It can be seen from Figure~\ref{Fig6}a that the trajectory of $x_2(t)$ is now noisy and  $L_2$ is again ahead of $L_1$; similar to Figure~\ref{Fig5}a. However, $L_1$ and $L_2$ are not synchronized. Since they are not synchronized, the peak of the cross-TLMI is then related to how well the two systems are correlated. This later correlation is related to $k_2$ when $\tau$ is fixed as in our case. When $k_2$ is too small, there is hardly any interaction and therefore very small correlation. However, when $k_2$ is too large, $L_2$ over-reacts to the influence from $L_1$ through the anticipatory term and the system becomes unstable with oscillatory responses. Therefore, there is an optimal $k_2$ for a given $\tau$. 
 
Figure~\ref{Fig6}c shows the history dependence of the TE for both directions. It can be see that their forms and values are similar to those from experiments. An important observation from Figure~\ref{Fig6} is that the DIF is correct only when the history length used for TE is below certain value; similar to experiments. Presumably, the effects of noise accumulates over time and information between two time points is lost if they are too far apart and therefore only history length below a certain value is valid for the computation of TE. The values of $k_2$ and $\lambda_2$ reported in Figure~\ref{Fig6} are chosen to produce TLMI and TE similar to those found in experiments.

\begin{figure}[h!]
 	\begin{center}
 		\includegraphics[width=8cm]{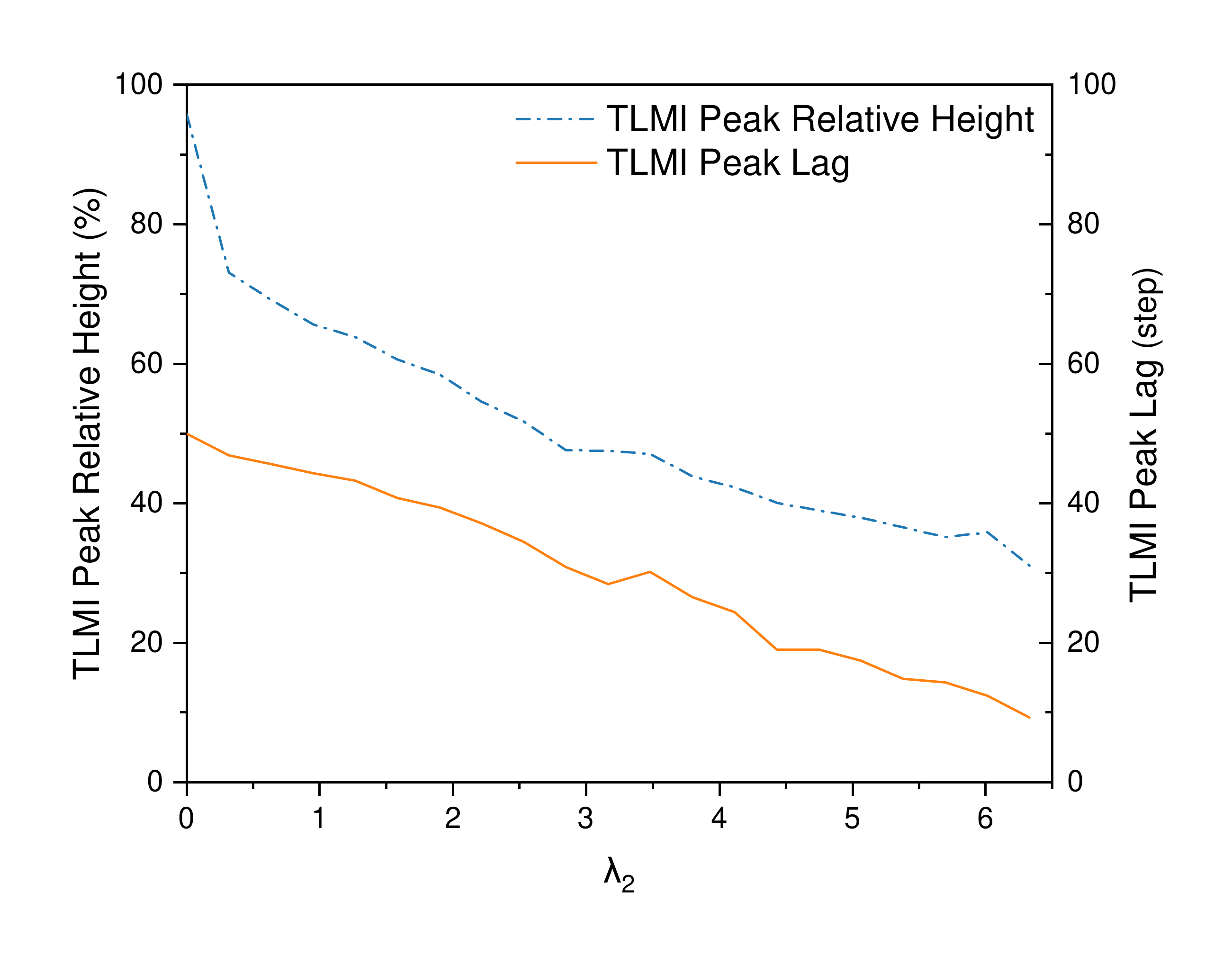}
 	\end{center}
 	\caption{Effects of noise on the anticipatory capability of $L_2$. The $\lambda_2$ dependence of height and location of the peak in the cross-TLMI for a simulation similar to that shown in Figure~\ref{Fig6}.}
 	\label{Fig7}
 \end{figure}

Since the synchronization (perfect correlation) is destroyed by the addition of noise, we found that both the height and the position of the peak of the cross-TLMI are function of noise strength $\lambda_2$ as shown in Figure~\ref{Fig7} which indicates that when the noise strength increases both the peak height of the cross-TLMI and the anticipatory horizon decrease. Of course, the noise amplitude cannot be too large; otherwise there will be no anticipation. When the noise amplitude is small, the shape of the attractor of $L_2$, although noisy, still resembles the Lorentz form of those from $L_1$. In such a case, there is still anticipation. However, when the noise amplitude is too large, the shape of the attractor of $L_2$ loses its Lorentz shape and there is no longer anticipation. We found that the effects of noise described above is not sensitive to where it is added. It can be added to the master or slave. Even when noises are added to both the master and slave, similar effects can be found.
\subsection{Mutual Anticipation}
\begin{figure}[h!]
 	\begin{center}
 		\includegraphics[width=5.4cm]{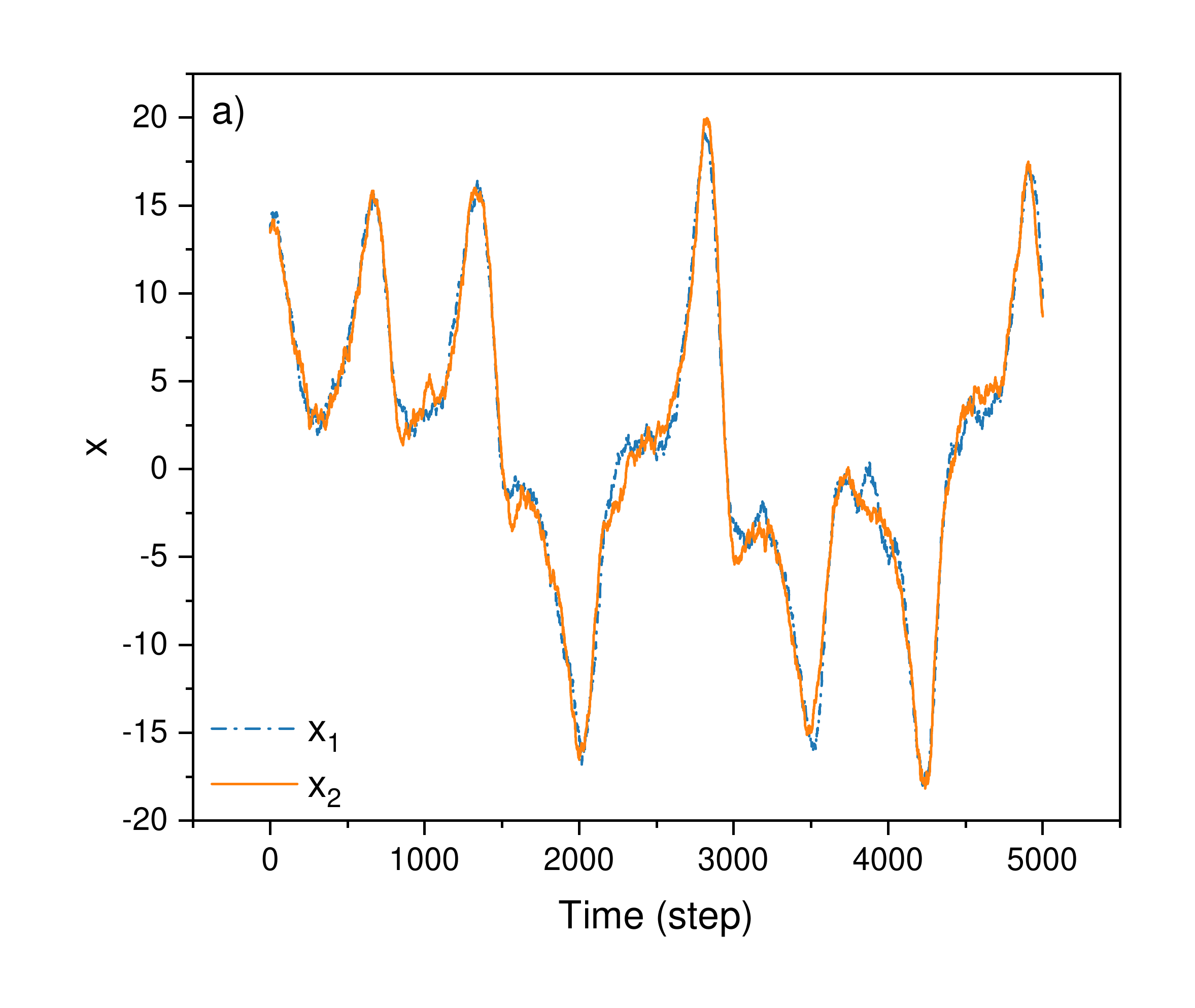}
 		\includegraphics[width=5.4cm]{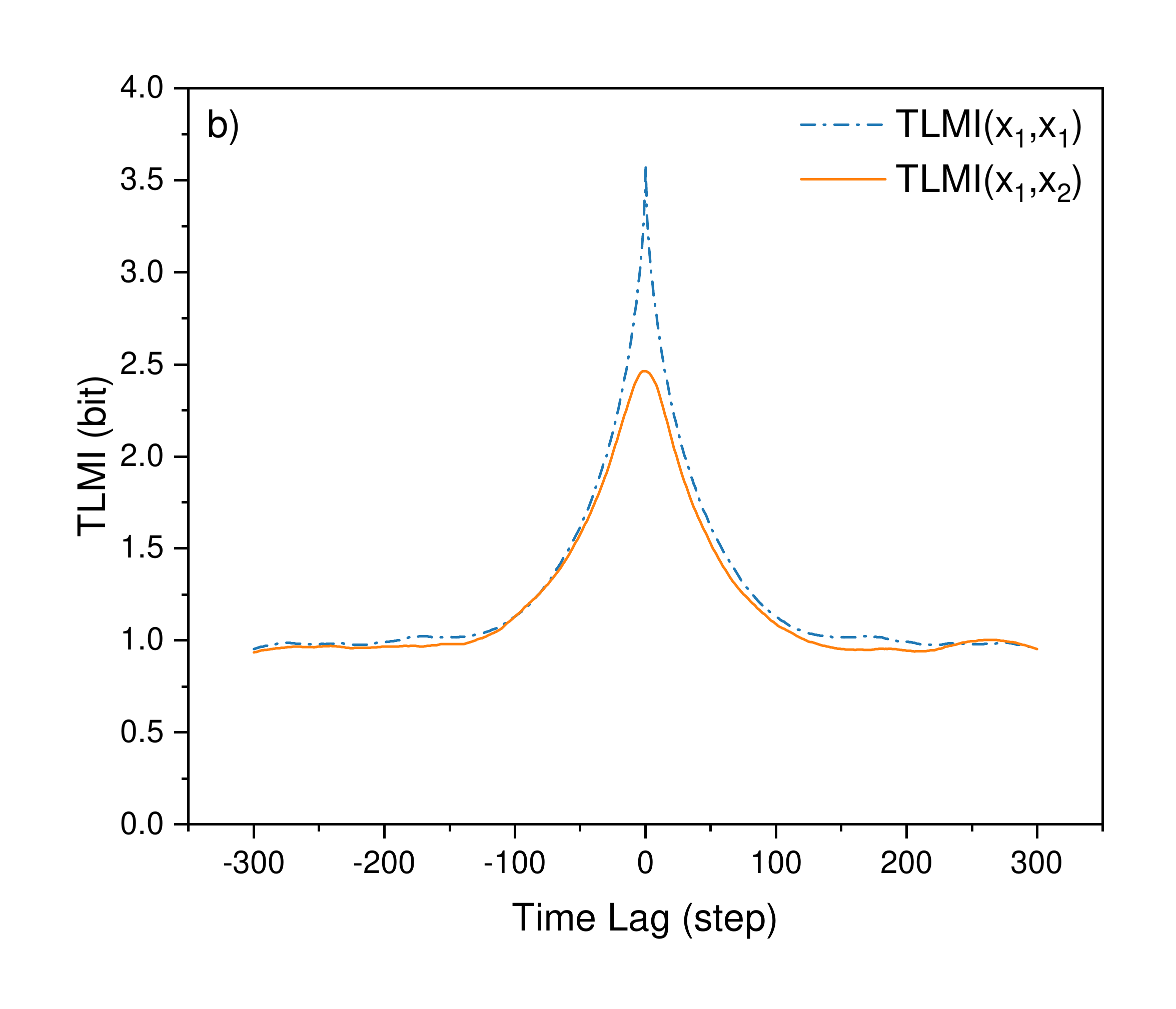}
 		\includegraphics[width=5.4cm]{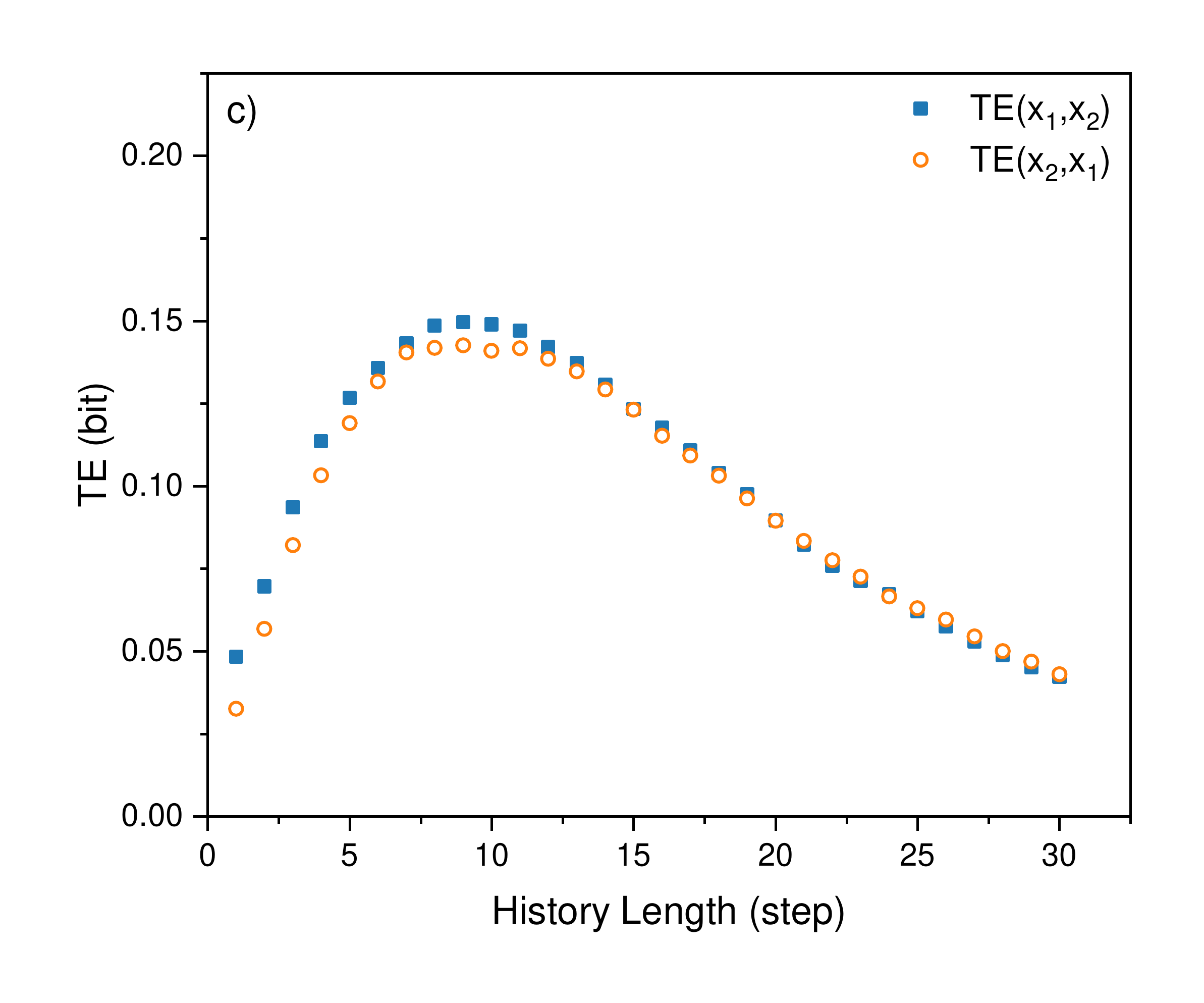}
 	\end{center}
 	\caption{Mutual anticipation between $L_1$ and $L_2$ with equal strength: a) The time course of the $x_1$ and $x_2$ of $L_1$ and $L_2$; showing no leader.  b) The TLMI between  $x_1$ and $x_2$ with peak close to the origin. c) A typical example of history length dependence of TE from $x_1$ to $x_2$ and vice versa; giving a difference of TE in these two directions very close but not equal to zero. The simulation parameters are: $\tau = 0.05$, $k_1=1, k_2=1, \lambda_1=3.8$ and $\lambda_2=3.8$.}
 	\label{Fig8}
 \end{figure}

\begin{figure}[h!]
 	\begin{center}
 		\includegraphics[width=5.4cm]{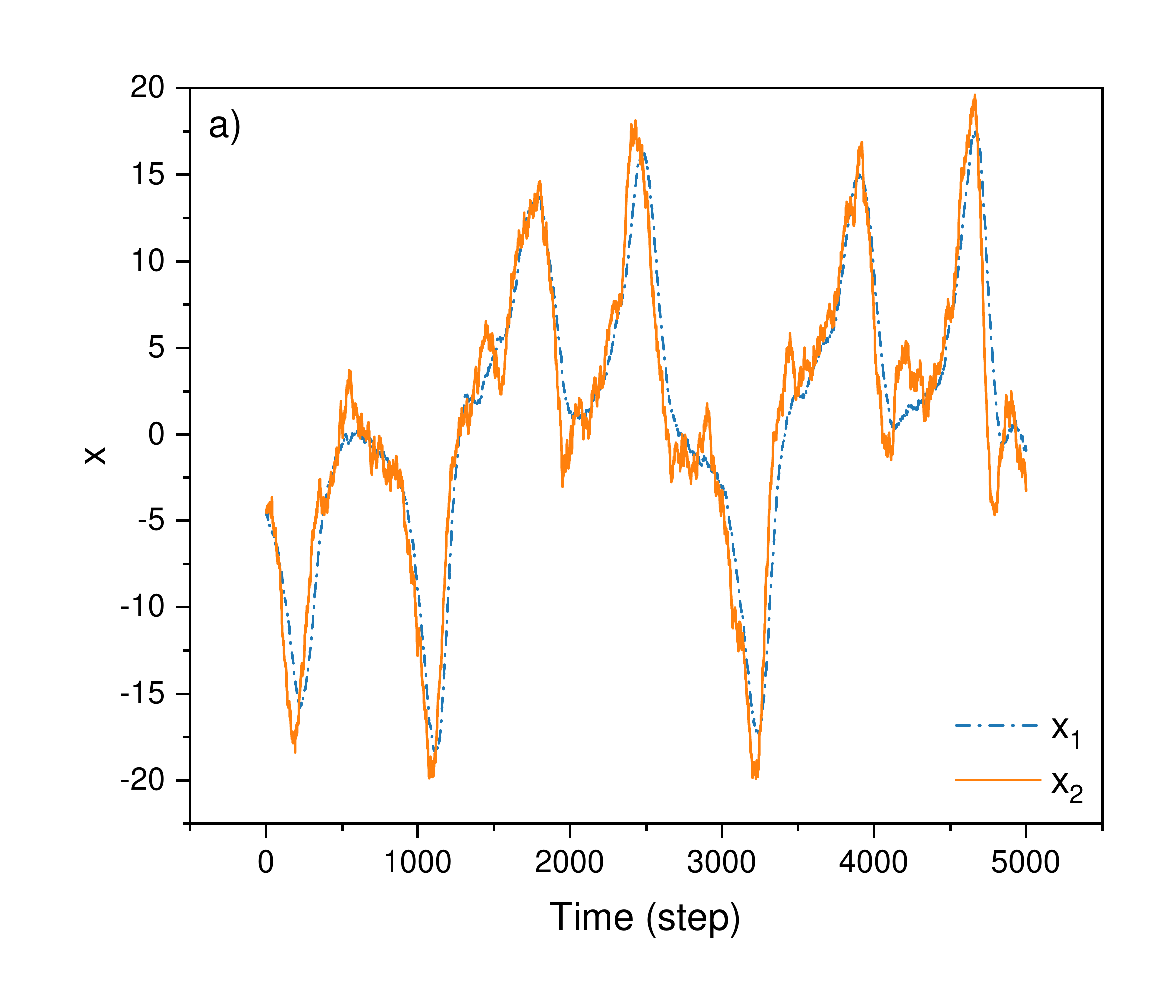}
 		\includegraphics[width=5.4cm]{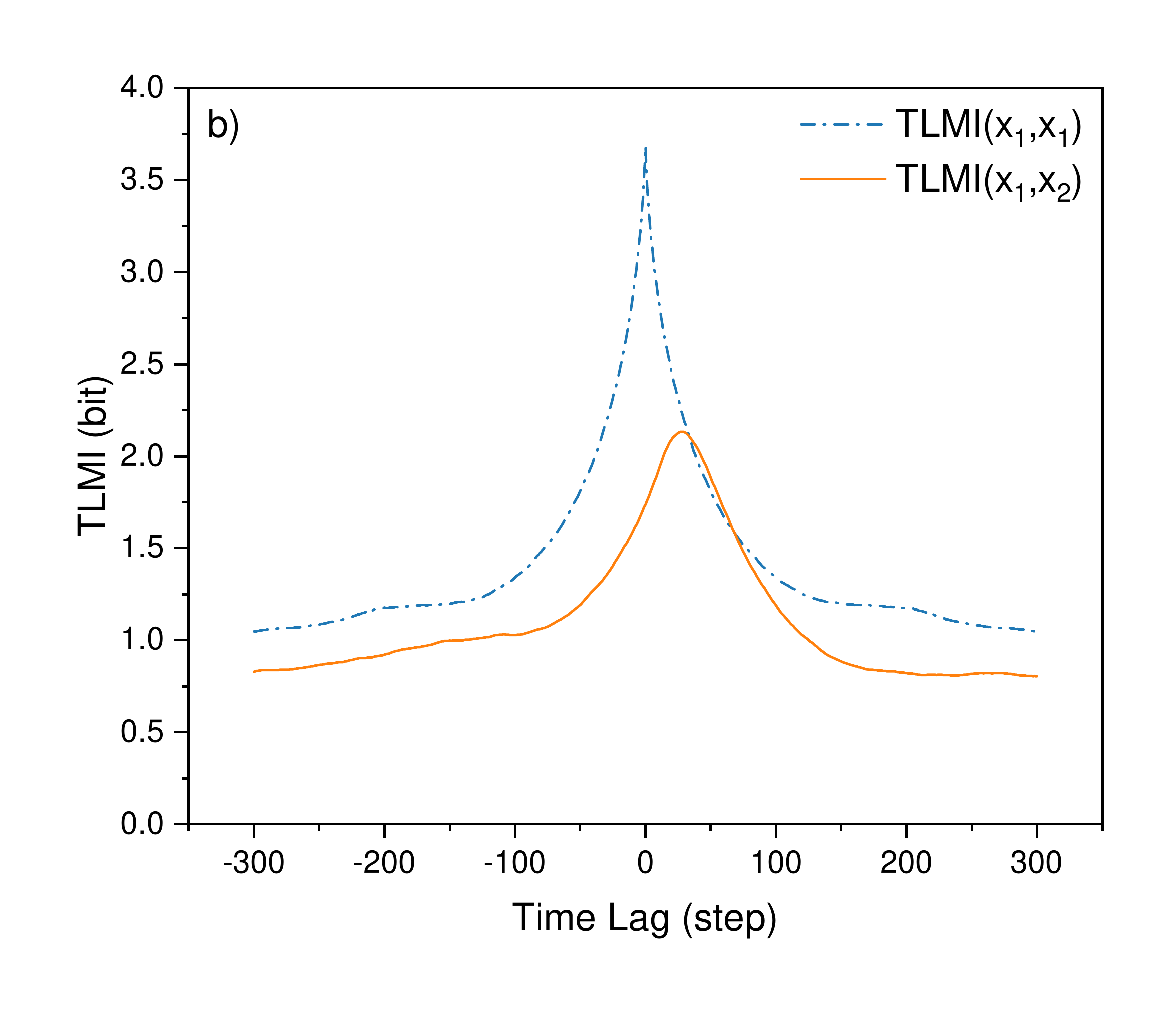}
 		\includegraphics[width=5.4cm]{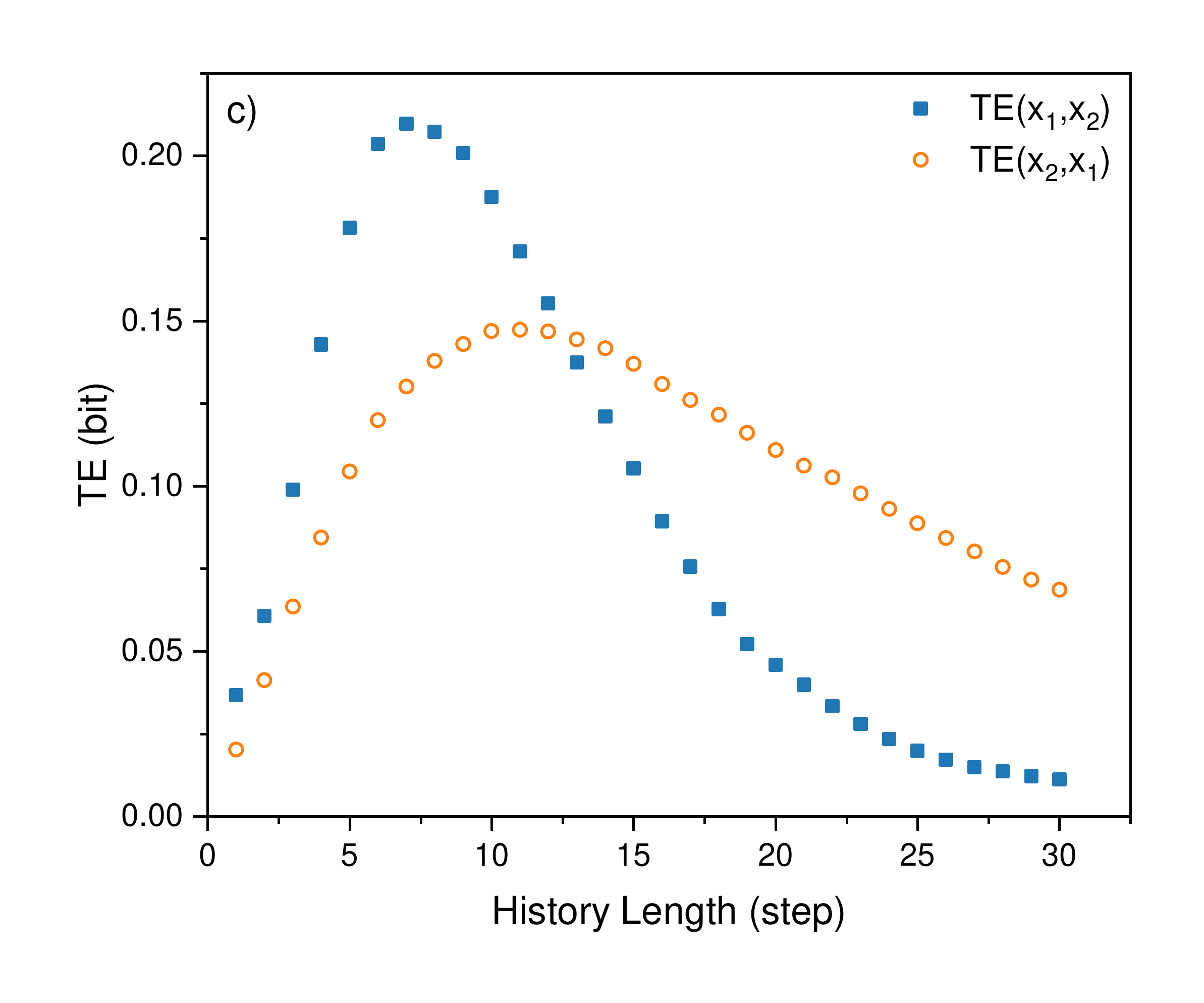}
 	\end{center}
 	\caption{Mutual anticipation between $L_1$ and $L_2$ with unequal strength and noises. a) The time course of the $x_1$ and $x_2$; showing the leading of $L_2$.  b) The TLMI between  $x_1$ and $x_2$. c) The history length dependence of TE from $x_1$ and $x_2$ and vice versa. The simulation parameters are: $\tau = 0.05$, $k_1=0.5, k_2=2.0, \lambda_1=3.8$ and $\lambda_2=3.8$.}
 	\label{Fig9}
 \end{figure}

For the case with both $k_1$ and $k_2$ being non-zero, we have the situation in which both systems are anticipatory of each other. This could very well be the case when two zebrafish meet and they anticipate each other's action. Figure~\ref{Fig8} is the typical results when $k1 = k_2$ with noises in both systems. From the figure, it can be seen that the peak of the cross-TLMI is close to the origin and there is no apparent leader. This is expected because of the symmetry of the interaction. However, because of noises, the positions of the peaks of these cross-TLMI from different realizations of the simulation can fluctuate but they are always close to the origin ($\pm 4$ step). Since the anticipation is symmetric, one would expect that the net information flow should be zero. However, it can be seen from Figure~\ref{Fig8}c that the net information transfer is close to but not equal to zero. Presumably, this is due to the stochastic nature of the dynamics. Different realizations will also give results similar to that shown in Figure~\ref{Fig8}c but with the DIF in the reverse direction. Therefore, no significant anticipatory dynamics is observed for such a symmetric case as expected. In order to produce anticipation, we found that there must be some significant difference between $k_1$ and $k_2$ as shown in Figure~\ref{Fig9}. Here, $k_2 = 4k_1$ and noises are needed to ensure the resemblance of findings from experiments. It can be seen from the figure that in such a case, the results are similar to those of a master-slave ($k_1 =0$) system. DIF can be correctly inferred from Figure~\ref{Fig9} when a history of less than certain steps is used; similar to the findings of experiments. Intuitively, this indicates that when one of the systems is dominating (with a smaller $k$), the situation is similar to that from a master-slave system.

\subsection{Anticipation by adaptation}
The anticipatory dynamics of the two Lorenz oscillators discussed above is the interaction of two systems with similar internal dynamics. We will call this anticipation as active. That is: $L_2$ is using its internal model (the same Lorenz dynamics) to perform anticipation of $L_1$. However, in the case of a passive anticipation, the anticipating system (slave) does not need to share similar internal dynamics or the ability to predict the internal dynamics of the master in order to anticipate. But it requires $L_2$ to have the ability to adapt to the dynamics of the master. For this case, we can use the relaxation dynamics introduced by Voss \cite{voss2016signal} to create an adaptive slave $L_3$ as:
\begin{eqnarray}
	{dx_3\over{dt}} = -\alpha x_3 +k_3[x_1(t)-x_3(t-\tau)] +\lambda_3\eta_3(t))
\end{eqnarray}
\noindent
Note that $L_3$ has only relaxation dynamics to ensure close to zero output when there is no input from the anticipatory term and does not require the knowledge of Lorenz dynamics to perform anticipation. The parameter $\alpha$ and $k_3$ are the relaxation and strength of anticipatory interaction of the system while $\lambda_3$ and $\eta_3$ are defined similarly as other $\lambda$ and $\eta$ above. It is through the anticipatory term from which $L_3$ can acquire (adapt to) the dynamics of $L_1$ when $\alpha$ and $k_3$ are properly chosen. Intuitively, one needs a sufficiently large $k_3$ to produce reasonable anticipation. However, from the simulation, it is found that in order to produce stable response from $L_3$, $k_3$ must not be too large and it is related to $\alpha$ for a fixed $\tau$. If $k_3$ is large, there will be oscillations in the response because of too much gain and then a larger $\alpha$ can be used to damp out the oscillations. Therefore, in order to generate stable response from $L_3$, $\alpha$ and $k_3$ cannot be varied independently. The $\alpha$ and $k_3$ reported here are chosen by using parameters scanning to produce TLMI and TE similar to those obtained from the experiments.

\begin{figure}[h!]
	\begin{center}
	    \includegraphics[width=5.4cm]{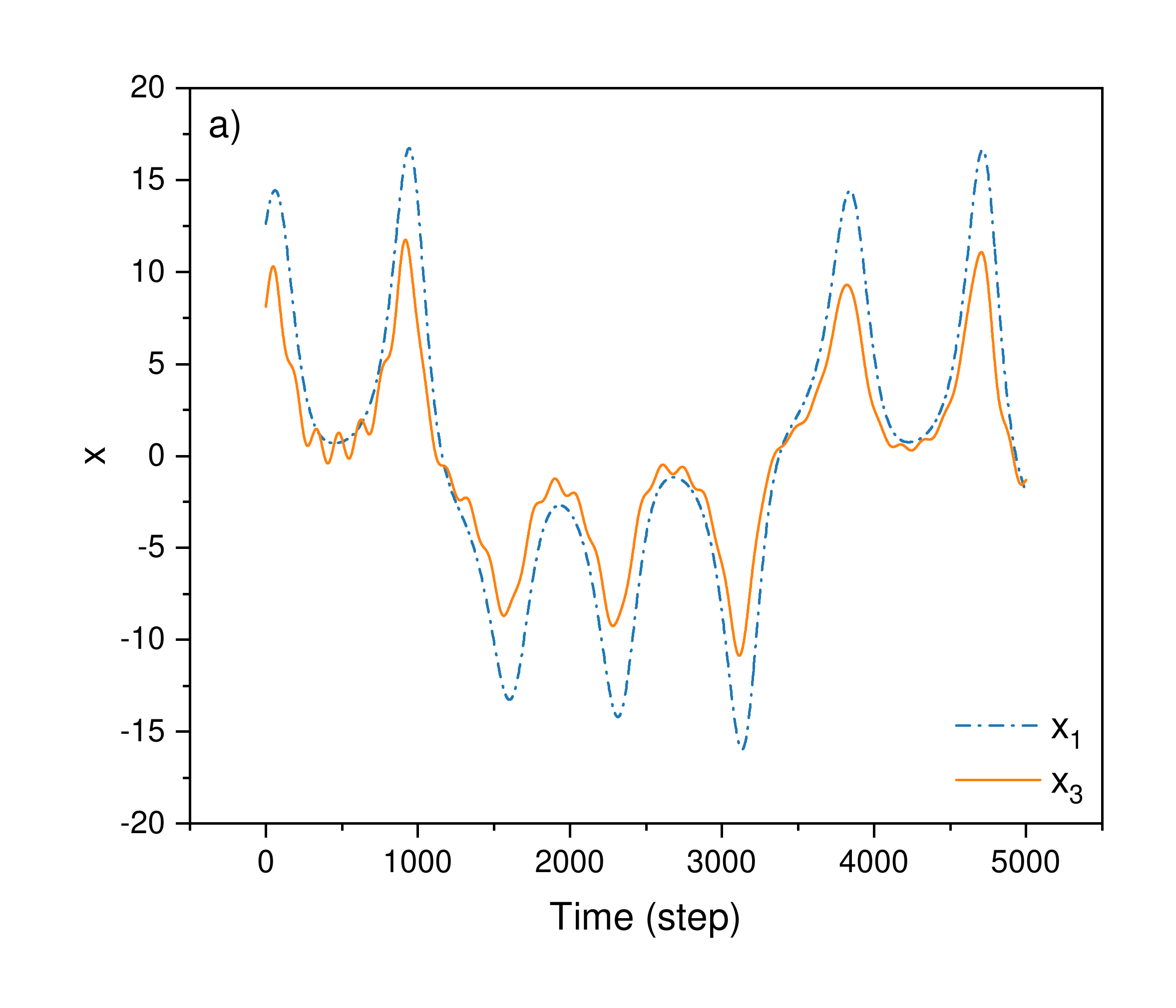}
		\includegraphics[width=5.4cm]{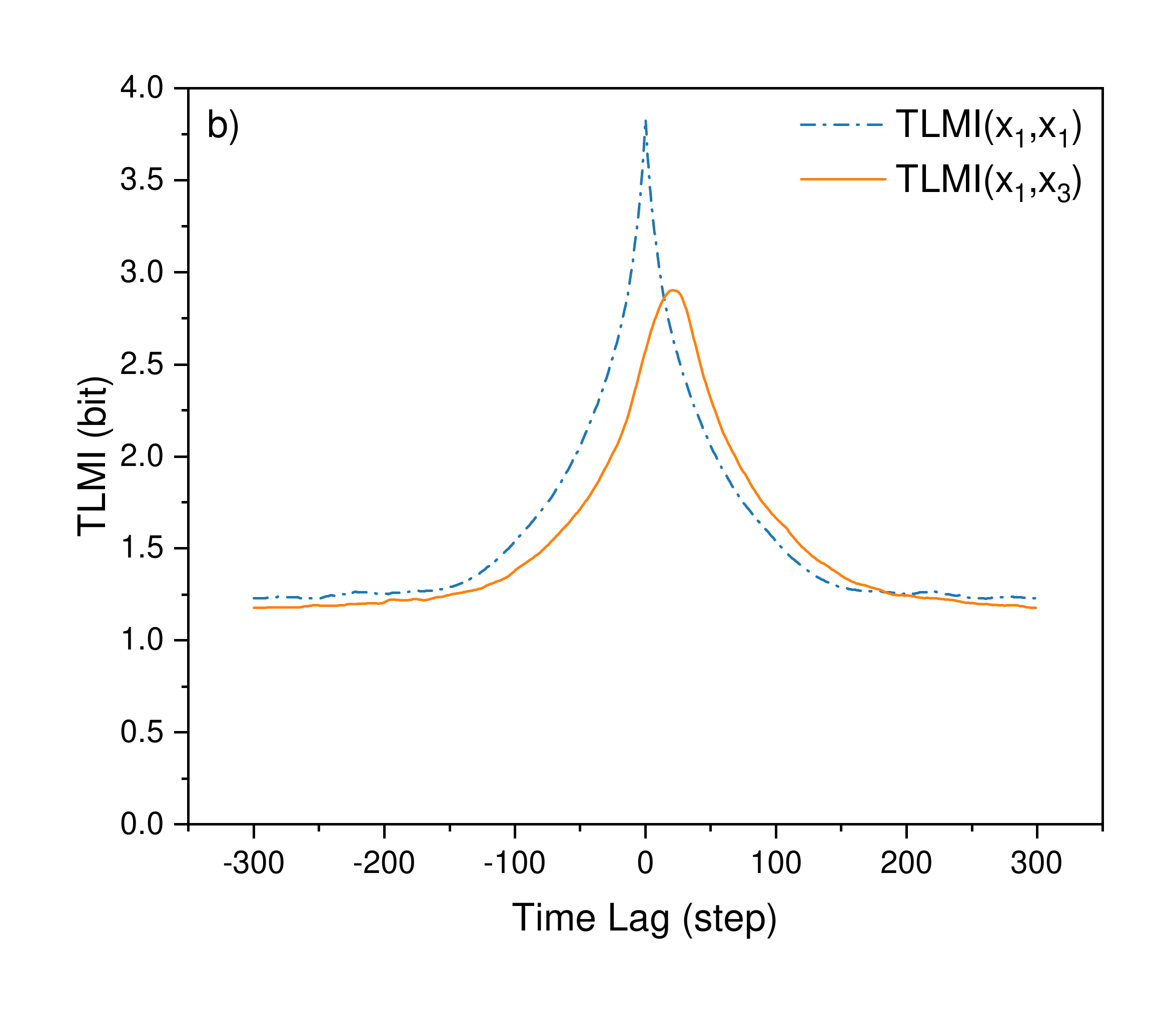}
		\includegraphics[width=5.4cm]{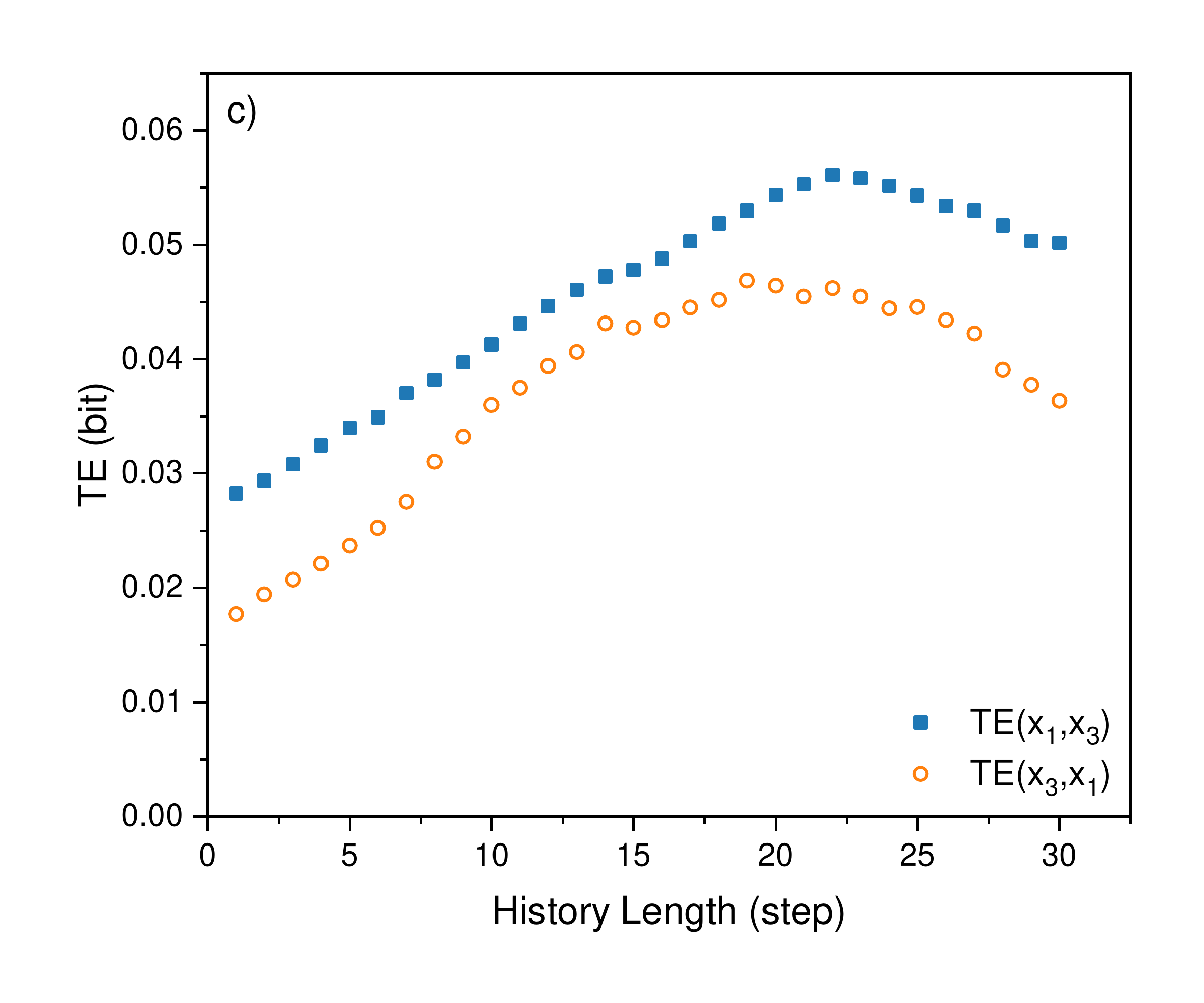}
	\end{center}
    \caption{Anticipation of $L_1$  using $L_3$ without noise. a) The time course of the $x_1$ and $x_3$ of $L_1$ and $L_3$; showing the leading of $L_3$.  b) The TLMI between  $x_1$ and $x_3$. c) The history length dependence of TE from $x_1$ to $x_3$ and vice versa. The direction of information flow is correct irrespective to history length. The simulation parameters are: $\tau = 0.05$, $k_1=0, k_3=55.0, \alpha = 34.0, \lambda_1=0$ and $\lambda_3=0$.}
	\label{Fig10}
\end{figure}

Figure~\ref{Fig10} shows a typical simulation of $L_1$ and $L_3$ without noise but with the properly chosen $k_3$ and $\alpha$ to mimic the experiment findings. Similar to the finding of Ref\cite{voss2016signal}, it can be seen that $x_3$ is anticipatory (ahead) of $x_1$. Since $L_3$ does not have the  dynamics of $L_1$, the cross-TLMI will always have a peak with height smaller than that of the auto-TLMI; no noises are needed. For the computed TE, Figure~\ref{Fig10}c shows that DIF is correct but its form is quite different from what we found in experiments. 

When moderate amount of noise is added to $L_1$ or $L_3$, or even to both $L_1$ and $L_3$ simultaneously, the effects of noise is to reduce both the height and the anticipatory horizon of the TLMI peak in Figure~\ref{Fig10}b; similar to those shown in Figure~\ref{Fig7}. However, the form of the history dependence of TE remains similar to Figure~\ref{Fig10}c. This last finding suggests that the form of the history dependence of TE is different for active and passive anticipation at least for the Lorenz system.

\begin{figure}[h!]
	\begin{center}
	    \includegraphics[width=5.4cm]{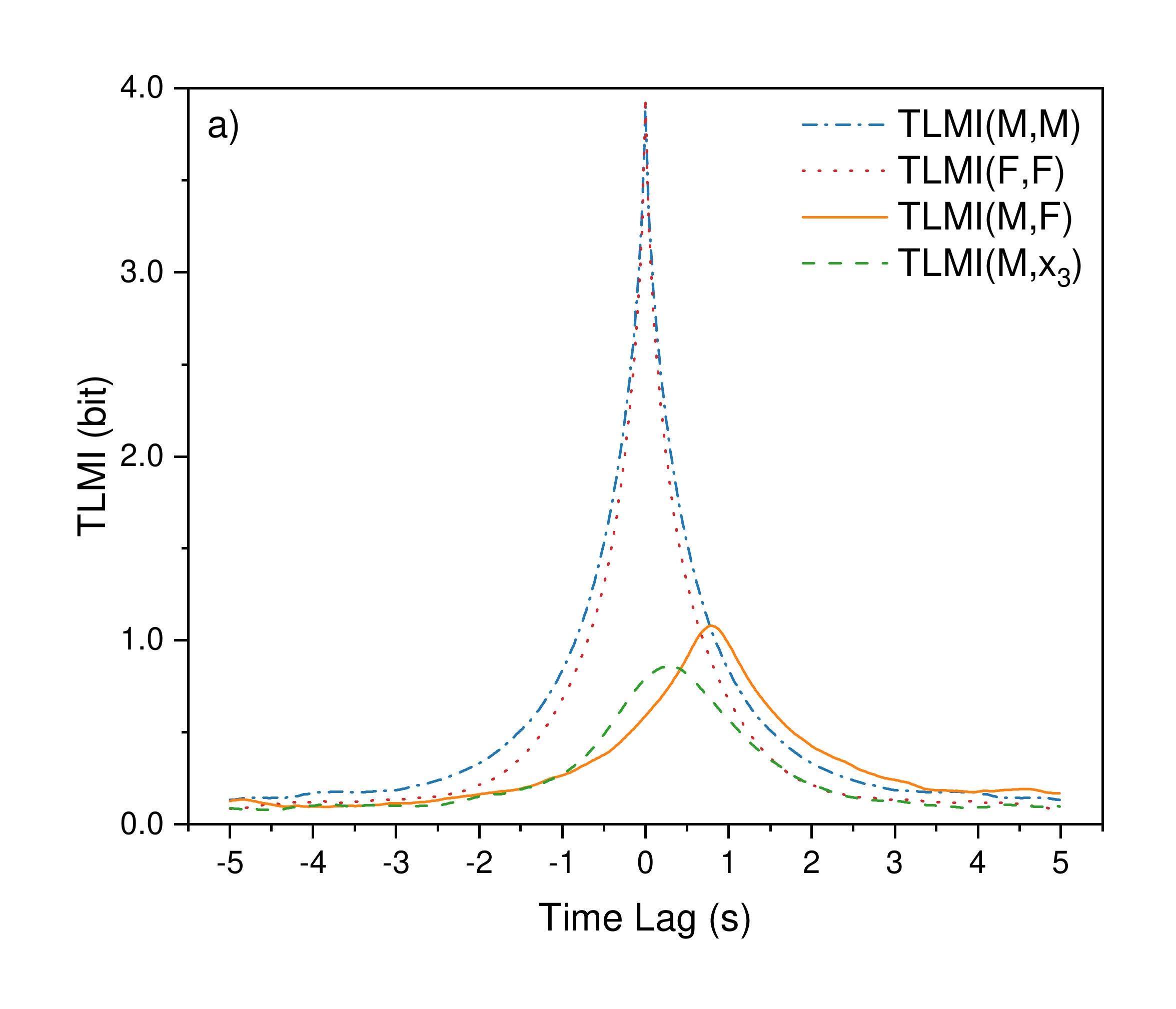}
		\includegraphics[width=5.4cm]{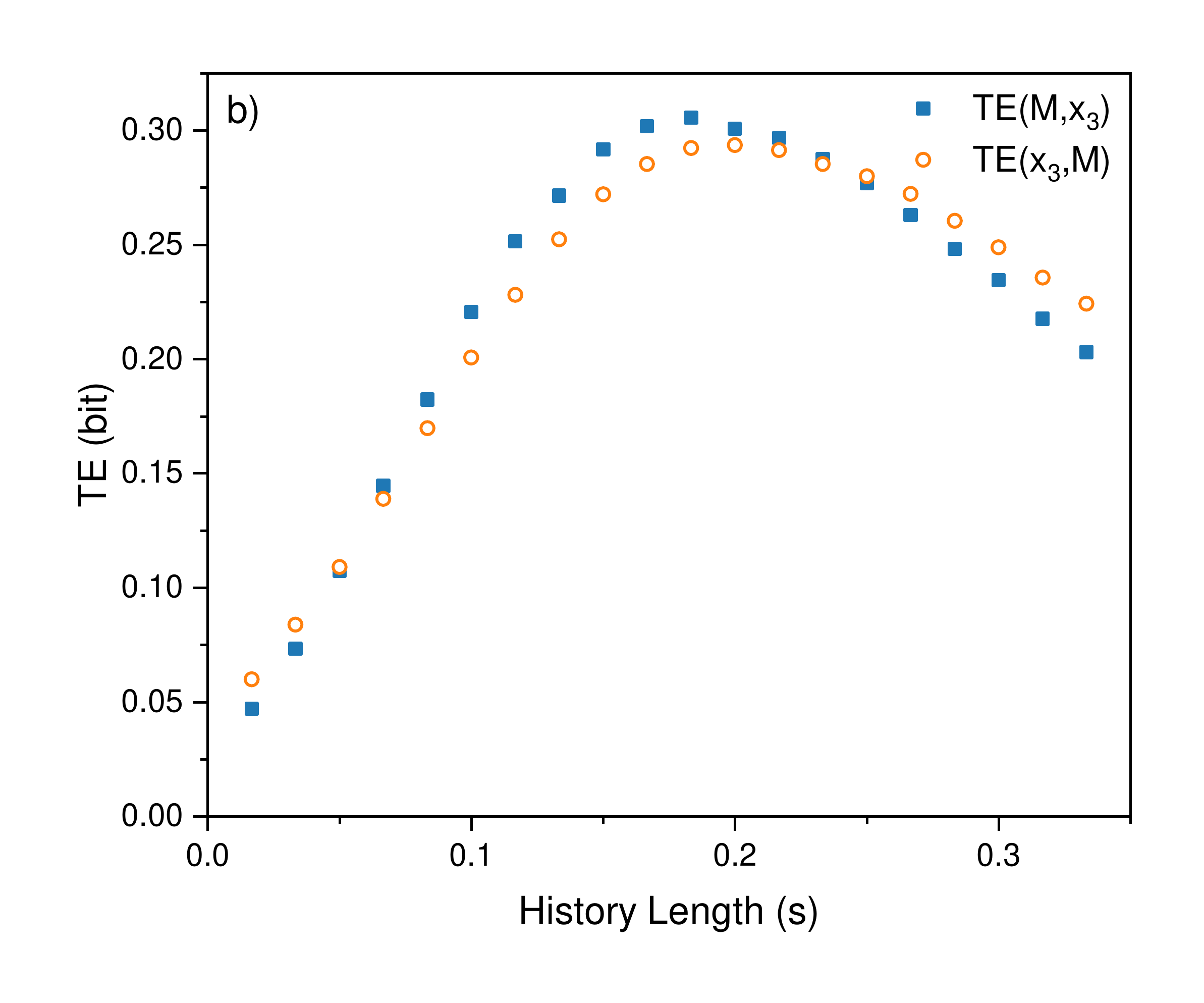}
	\end{center}
    \caption{Anticipation of experimental data from the experiment shown in Figure~\ref{Fig2} using $L_3$. a) Same as Figure~\ref{Fig1}a with the addition of cross-TLMI between L3 and experimental data for comparison. b) History dependence of TE in both directions. The simulation parameters for $L_3$ are: $\tau = 0.05$, $k_3=150.0, \alpha = 150.0$ and $\lambda_3=0$. To match the simulation with the experiment, one simulation time step ($\Delta t = 0.001$) corresponds to 1/fps in the experiment which is 16.6 ms.}
	\label{Fig11}
\end{figure}

Finally, we can also use $L_3$ to perform anticipation on the trajectory of a master obtained from an experiment with anticipation as shown in Figure~\ref{Fig11} to see if passive anticipation is taking place during fish interaction. In Figure~\ref{Fig11}, one simulation step is set to correspond to 16.6 ms (1/fps) in our experiments and $x_1(t)$ in Eqn(7) is replaced by the trajectory of a master. Also, we have done a parameter scan to obtain the best anticipatory capability for $L_3$. It can be seen from the figure that although the direction of information flow is correct (Figure~\ref{Fig11}b) for history length between 67 and 233 ms, the anticipatory capability of $x_3$ is not as good as that obtained from a real fish! Both the peak height and the anticipatory horizon of the cross-TLMI are smaller from those obtained from experimental results. Figure~\ref{Fig11} suggests that the fish are using their own similar internal dynamics for anticipation


.

\section{Discussions}

When Wiener first proposed the idea of using the history of one time series to help for the prediction for the future values of another time series, he had in mind that causal events should be ordered by time \cite{beckenbach2013modern}. However, as it can be seen above that when anticipation dynamics are involved, causal events need not be ordered by time. This happens when the interacting agents are active in the sense that its dynamics is affected by the incoming information. This aspect of anticipatory dynamics might be a new paradigm for interaction between biological entities.


One interesting aspect of anticipatory dynamics is that the leading in time course does not necessary means the "leader" \cite{nagy2010hierarchical} is in control. In the phenomenon of anticipating synchronization demonstrated in Ref\cite{voss2000anticipating}, it can be clearly seen that the "leader" (time course) can be the slave because the slave is actively anticipating the future movements of the master. It is known that when birds flock, the birds from low pecking order can lead (in the sense of time course) for sometimes. Perhaps, these low pecking order birds are just anticipating the future position/movement of their masters (birds of high pecking order). We are still in the early stage in understanding these emerging behavior from these active particles in the sense they are not only self-propelling but at the same time making active decision.

Our method of using TLMI and TE to detect causality in anticipatory system works only when history length used in TE can be empirically calibrated by the one-way mirror experiments. There is still not a clear criterion of correct history length. Intuitively, the correct history length should be related to the anticipation dynamics of the system. It is still not clear why one should be using a history length with a upper critical value as in the case of zebrafish and Lorenz oscillators studied here. Also, the interpretation of the 2-peak TLMI is still difficult. We were not able to reproduce the double-peaked TLMI with reason parameters by using the Lorenz oscillators. Presumably, more complex dynamics are involved for these double-peaked TLMI. Finally, our distinction between active and passive anticipation can be tested further by studying interaction between preys and predators in fish. Presumably, predators and preys are of different species and therefore prey and predators should be using passive anticipation to avoid or pursuit their targets. It would be interesting to find out if predators can learn from their experience to change its anticipatory dynamics from a passive to an active one.

One of the author (CKC) would like to thank Prof. Viscek for inspiring discussions. This work has been supported by the MOST of ROC under the grant number 109-2927-I-001-507.

\bibliographystyle{prsty}
\bibliographystyle{prsty}
\bibliography{references}



\end{document}